\documentclass[structabstract]{aanew}  
\usepackage{graphicx}
\usepackage{natbib}
\bibpunct{(}{)}{;}{a}{}{,} 
\usepackage{txfonts}
\begin{document}
   \title{Search for Extratidal Features Around 17 Globular Clusters in the Sloan Digital Sky Survey}


   \author{Katrin Jordi
          \inst{1,2}
          \and
          Eva~K. Grebel\inst{1}
	  }

   \institute{Astronomisches Rechen-Institut, Zentrum f\"ur Astronomie der Universit\"at Heidelberg, M\"onchhofstrasse 12-14, 69120 Heidelberg, Germany
     \and 
     Department of Physics, University of Basel, Klingelbergstrasse 82, 4056 Basel, Switzerland \\\email{jordi@ari.uni-heidelberg.de}
             }

   \date{Received March 2010; accepted August 2010}

 \abstract{}{}{}{}{} 
 
  \abstract
   {The dynamical evolution of a single globular cluster and also of the entire Galactic globular cluster system has been studied theoretically in detail. In particular, simulations show how the `lost' stars are distributed in tidal tails emerging from the clusters. }
   {We investigate the distribution of Galactic globular cluster stars on the sky to identify such features like tidal tails. The Sloan Digital Sky Survey provides consistent photometry of a large part of the sky to study the projected two dimensional structure of the 17 globular clusters in its survey area.}
   {We use a color-magnitude weighted counting algorithm to map (potential) cluster member stars on the sky. The same algorithm was already used by Odenkirchen et al. (2003) for the detection of the tidal tails of Palomar~5. }
   {We recover the already known tidal tails of Pal~5 and NGC~5466. For NGC~4147 we have found a two arm morphology. Possible indications of tidal tails are also seen around NGC~5053 and NGC~7078, supporting earlier suggestions. Moreover, we find potential tails around NGC~5904 and Pal~14. Especially for the Palomar clusters other than Pal~5, deeper data are needed in order to confirm or to rule out the existence of tails. For many of the remaining clusters in our sample we observe a pronounced extratidal halo, which is particularly large for NGC~7006 and Pal~1. In some cases, the extratidal halos may be associated with the stream of the Sagittarius dwarf spheroidal galaxy (e.g.,NGC~4147, NGC~5024, NGC~5053).}
   {}


   \maketitle
%

\section{Introduction}
Globular clusters (GCs) are the oldest stellar objects found in the Milky Way (MW), with ages in the range of $10-13$~Gyr. They represent tracers of the early formation history of our Galaxy. GCs are ideal systems to study stellar dynamics in high-density systems as their relaxation times are much smaller than their age \citep[GO97]{gnedin97}. We expect, at least in the cluster core, the stars to have lost memory about their initial conditions. The GCs we see today are only the lucky survivors of an initially much larger population (GO97). Several different mechanisms can dissolve a GC: internal processes include infant mortality \citep{geyer01,kroupa01} by stellar evolution, gas expulsion and two-body relaxation; external mechanisms include primarily the interaction of the GC with the constantly changing tidal field of its host galaxy. Different processes are dominant at different ages of the cluster. For young clusters stellar evolution is the main driver that may potentially cause them to dissolve. More massive stars are evolving rapidly, have strong winds and explode as supernovae within the first few million years driving out the remaining gas. The sudden change, due to the mass loss, in the GC's internal gravitational field can destroy the GC. Two-body relaxation arises from close encounters between cluster stars, leading to a slow diffusion of stars over the tidal boundary. This process is taking place from the beginning to the final dissolution of the cluster. Equipartition of energy by two-body encounters leads to mass-segregation, the more massive stars sink to the center, less massive stars remain in the outer parts of the cluster. This leads to a preferential loss of low-mass stars, which can be observed today in declining mass-function slopes of GCs  \citep{baumgardt03}. On the other side, the mass loss due to the cluster's encounter with the Galactic disk and bulge is strongest at pericenter passages, especially for GCs on elliptical orbits \citep{baumgardt03}. All these mechanisms have one common result: the GC is losing stars. As a result of relaxation stars gradually diffuse out, while during tidal shocks the GC's loss of stars is more sudden. 

\subsection{Observational studies}
\citet{grillmair95} performed deep two-color star counts to examine the outer structure of 12 GCs using photographic Schmidt plates. They detected a halo of extratidal stars around  most of their clusters. They also found indications of possible tidal tails emerging from some clusters. The tidal tails are not in all cases purely of stellar origin, as the morphological identification of galaxies and stars, especially for faint objects, was not perfect. I.e., in some cases background galaxies were mistaken for stars. In other cases random overdensities of foreground stars induced features into the cluster's contours. \citet{leon00} studied the tidal tails of 20 GCs, using Schmidt plates as well. These authors also find halos of extratidal stars as well as tidal features for most of their clusters. The two studies have three GCs in common. For two out of the three clusters \citet{grillmair95} detected tidal tails where \citet{leon00} did not see such a strong signal and in the third case it is the opposite.

Apart from the two studies mentioned above, different authors determined the 2d distribution of cluster stars on the sky for individual clusters. The two most prominent cases are Pal~5 and NGC~5466, as for these two clusters extended tidal tails have been found. The large area-coverage with CCDs by the SDSS provides a large database for tidal tail studies of GCs. Pal~5 was the first cluster for which extended tidal tails were discovered and studied in detail \citep{odenkirchen01,rockosi02,odenkirchen03,grillmair06,odenkirchen09} using SDSS data. NGC~5466 was the second cluster found to have extended tidal tails \citep{odenkirchen04,belokurov06,grillmair06_5466}. Not only spectacular tidal tails, but simpler tidal features were also detected for NGC~5053 \citep{lauchner06}, for NGC~6341 (M92) by \citet{testa00}, and for NGC~4147 by \citet{martinez04}. Further, \citet{kiss07} studied the kinematics of red giants around five GCs. For M55, M30 and NGC~288, they did not find any strong signs of tidal debris. Recently, \citet{chun10} studied the spatial configuration of stars around five metal-poor GCs (M15, M30, NGC~5024 (M53), NGC~5053, and NGC~5466). They used deep images obtained at the Canada-France-Hawaii Telescope. Around all GCs extratidal overdensities and extratidal halos were observed. Between NGC~5024 and NGC~5053 they detect a tidal-bridge like feature and an envelope structure.

\subsection{Theoretical prediction and N-body simulations}
\citet{gnedin97} (GO97) studied the dynamical evolution of the entire Galactic GC system. They derived the destruction rates due to the internal and external mechanisms. According to their calculations, the total destruction rate of the entire GC system is such that more than half of the present day GCs will not survive the next Hubble time. GO97 did not include any information on the GCs' proper motion, they used the known present day distance to the Galactic center and the GCs' radial velocity.

\begin{figure}
\centering
\includegraphics[width=9cm]{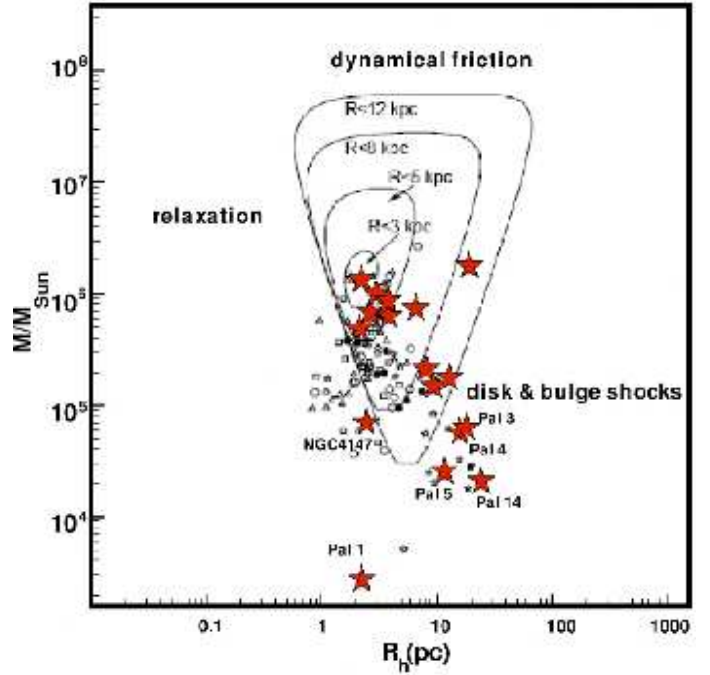}
\caption{Vital diagram from GO97. The red stars are the GCs in our sample, Cluster mass and half-light radius were taken from GO97. GCs within the triangles are most likely surviving the next Hubble time. Adopted from G097, reproduced by permission of the AAS}
\label{figVital}
\end{figure}

\citet{dinescu99} (DG99) derived the destruction rates for a smaller sample of Galactic GCs in a similar way, but included the measured proper motion data. They concluded that the destruction processes for the clusters in their sample are mostly dominated by internal relaxation and stellar evaporation. Tidal shocks due to the bulge and disk are only dominant for a small number of GCs in their sample. For the sample we are studying here this is only the case for Pal~5. Comparing their destruction rates with the numbers of GO97, they find that the GO97 total rates are larger. The destruction rates due to internal processes (2-body relaxation, stellar evaporation) are comparable in both studies, therefore only the destruction rate due to tidal shocks is incompatible. DG99 conclude that the statistical approach of assigning tangential velocities (the method used by GO97) results in more destructive orbits than are actually observed. 

A similar conclusion is drawn by \citet{allen06}, who derive the orbits of Galactic GCs, also based on observed proper motions, in an axisymmetrical potential and in a barred potential (resembling the MW). They derived the destruction rates for the GCs in their sample in an almost identical way to DG99. \citet{allen06} further investigated the influence of the bar on the clusters' orbits and destruction rates. The bar does not influence the orbits of clusters that have a pericenter distance greater than $\sim4$~kpc. Hence, none of the clusters in our sample (see Table~\ref{tblgcs}) are influenced by the bar.

\citet{combes99} studied the tidal effects experienced by GCs on their orbits around the MW. In general, they find that two large tidal tails should emerge extending out to $5$ tidal radii. They conclude by stating \emph{i)} GCs are always surrounded by tidal tails and tidal debris, \emph{ii)} the tails are preferentially composed of low mass stars, \emph{iii)} mass loss in a GC is enhanced if the cluster is rotating in the direct sense with respect to its orbit, \emph{iv)} extended tidal tails trace the cluster's orbit, and \emph{v)} stars are not distributed homogeneously along the tidal tails, but clump. 

\citet{capuzzo05} investigated the clumpy structure of the tidal tails. They found that the clumpy structure is not associated with an episodic mass loss or tidal shocks due to encounters with Galactic substructure, as stars are continuously lost. The clumps are not self-gravitating systems. \citet{kuepper08} showed that the clumps are a result of epicyclic motion of the stars lost by the cluster. \citet{just09} investigated the formation of the clumps and discussed their observability. \citet{kuepper09} investigate the position of the clumps relative to the cluster for clusters on eccentric orbits. They concluded that tidal shocks are not the direct origin of the clumps. Further, \citet{capuzzo05} confirmed the earlier finding that the cluster's leading tail develops inside the orbit, while the trailing tail follows outside the orbital path. They also found that for clusters on elliptical orbits the tidal tails trace the orbital path only near perigalacticon \citep[see also][]{dehnen04}. On the other hand for circular orbits the tails are clear tracers of the cluster path. Moreover the length of the tails is not constantly growing. On eccentric orbits the leading tail tends to be longer than the trailing tail during the cluster's motion from apogalacticon to perigalacticon and vice versa during the other half of the cluster's orbit.

\citet{montuori07} investigated the direction of tidal tails with respect to the GC's orbit. In the outer parts of the tails ($>7-8$ tidal radii away from the cluster center), the tails are very well aligned with the cluster's orbit regardless of the cluster's location on the orbit. On the other hand, in the inner parts, the orientation of the tidal tails is strongly correlated with the orbital eccentricity and the GC's location on the orbit. Only if the cluster is near perigalacticon, the inner tidal tails are aligned with the orbital path. Therefore, only if long tidal tails are detected, it is possible to securely constrain the cluster's orbit from those. Detecting only small, short tidal extensions just outside the GC's tidal radius do not give any hint on the cluster's orbit unless the cluster's proper motion has been measured before.

DG99 published proper motions of a large number of GCs and added values from the literature to their catalog. From this they derived orbital parameters, such as ellipticity $e$, perigalacticon $R_{peri}$, apogalacticon $R_{apo}$, etc. The eleven clusters we have in common with DG99 have orbital eccentricities larger than $\sim0.3$, i.e., there are no circular orbits for our sample.

These studies give a good theoretical understanding of the size, shape, orientation, etc. of tidal tails of GCs. In an observational study of light element abundance variations in field halo stars, \citet{martell10} conclude that as many as half of the stars in the Galactic halo may originally have formed in meanwhile dissolved GCs. In this paper we use data from the Sloan Digital Sky Survey to study the 2d structure of 17 Milky Way GCs and to compare the observed features with the theoretical predictions. The paper is organized as follows: Section~\ref{Secdata} introduces the photometric data used in our analysis. Section~\ref{Seccounting} explains the algorithm used to count (potential) member stars on the sky. In Section~\ref{Secprofiles} we present the number density profiles of the 17 clusters in our sample. Section~\ref{Sectidal} presents the derived contour maps for our GCs. In Section~\ref{Secdiscussion} we discuss our results. The paper is concluded in Section~\ref{Secsummary} with a summary.


\section{Data}
\label{Secdata}
The Sloan Digital Sky Survey (SDSS) is an imaging and spectroscopic survey in the Northern hemisphere \citep{york00}. SDSS imaging data are produced in the five bands \emph{ugriz} \citep{fukugita96,gunn98,hogg01,smith02,ivezic04,tucker06,gunn06}. The data are automatically processed to measure photometric and astrometric properties \citep[Photo]{lupton02,pier03} and are publicly available on the SDSS web pages\footnote{www.sdss.org}. We used the data from the SDSS Data Release 7 \citep[DR7;][]{abazajian09}.

The automatic SDSS pipeline Photo was initially designed to process high Galactic latitude fields with a low density of Galactic field stars. Fields centered on and around GCs are too crowded for Photo to process, so the automatic pipeline does not provide photometry for these most crowded regions. \citet[An08]{an08} used the DAOPHOT crowded-field photometry package to derive accurate photometry for the stars in these crowded areas. This photometry is published on the SDSS web pages as a value-added catalog\footnote{http://www.sdss.org/dr7/products/value\_added/index.html}. From this catalog we only considered stars with photometry flags set to 0. The An08 catalog and the original SDSS catalog overlap around each GC. If a star is measured in both samples, we used the original SDSS photometry. This choice has no influence on the result. Finally, we merged the two datasets into one catalog for each GCs. The magnitudes in the final catalog were corrected for extinction. The SDSS catalog provides the Galactic foreground extinction values from \citet{schlegel98}. For the fields with An08 photometry, the extinction values were derived by a cubic interpolation of the values listed in the SDSS catalog.

In Table~\ref{tblgcs} we list the GCs found in the SDSS DR7 footprint. Columns (3) and (4) contain the equatorial coordinates (J2000.0) of the cluster center mainly taken from \citet{harris96}, except for Pal~14 \citep{hilker06} and NGC~7089 \citep{dalessandro09}. In column (5) we list the clusters' distance to the Sun and in column (6) the clusters' distance to the center of the MW from \citet{harris96}, except for Pal~14 \citep{hilker06}. Columns (7) and (8) contain the core and tidal radii from \citet{mclaughlin05} (M05). For almost all GCs in our sample proper motions have been measured. Only for three of the most distant GCs, NGC~2419, Pal~4 and Pal 14, as well as for NGC~5053 this information is lacking. Columns (9) and (10) list the proper motion taken from DG99. 

\begin{table*}
\caption{Globular clusters in the SDSS DR7 footprint.\label{tblgcs}}
\centering
\begin{tabular}{llccrrrrcc}
\hline
\hline
\noalign{\smallskip} 
{NGC} & {Name} & {RA} & {Dec} & {$R_\odot$} &{$R_{MW}$} & $r_c$ & $r_t$ & $\mu_{\alpha}\cos(\delta)$ & $\mu_{\delta}$ \\
{} &  {} & {(hh mm ss.s)} & {($^\circ$ $\arcmin$ $\arcsec$)} & {(kpc)} & {(kpc)} & {(\arcmin)} & {(\arcmin)}&{(mas/yr)} & {(mas/yr)}\\
\noalign{\smallskip} 
\hline
\noalign{\smallskip} 
2419 &       & 07 38 08.5 & +38 52 55 & 84.2  & 91.5 &0.32 & 8.34 & \dots          &\dots           \\
4147 &       & 12 10 06.2 & +18 32 31 & 19.3  & 21.3 &0.09 & 6.32 & $-1.85\pm0.82$ & $-1.3\pm0.82$  \\
5024 & M~53  & 13 12 55.3 & +18 10 09 & 17.8  & 18.3 &0.35 &16.06 & $+0.5\pm1$     & $-0.1\pm1$     \\
5053 &       & 13 16 27.0 & +17 41 53 & 16.4  & 16.9 &1.91 &13.85 & \dots          & \dots          \\
5272 & M~3   & 13 42 11.2 & +28 22 32 & 10.4  & 12.2 &0.37 &30.15 & $-1.1\pm0.51$  & $-2.3\pm0.54$  \\
5466 &       & 14 05 27.3 & +28 32 04 & 15.9  & 16.2 &1.53 &13.64 & $-4.65\pm0.82$ & $0.80\pm0.82$  \\
5904 & M~5   & 15 18 33.8 & +02 04 58 & 7.5   &  6.2 &0.42 &22.45 & $+5.07\pm0.68$ & $-10.7\pm0.56$ \\
6205 & M~13  & 16 41 41.5 & +36 27 37 & 7.7   &  8.7 &0.55 &19.49 & $-0.9\pm0.71$  & $+5.5\pm1.12$  \\
6341 & M~92  & 17 17 07.3 & +43 08 11 & 8.2   &  9.6 &0.26 &12.96 & $-3.3\pm0.55$  & $-0.33\pm0.70$ \\
7006 &       & 21 01 29.5 & +16 11 15 & 41.5  & 38.8 &0.32 &12.75 & $-0.96\pm0.35$ & $-1.14\pm0.40$ \\
7078 & M~15  & 21 29 58.3 & +12 10 01 & 10.3  & 10.4 &0.07 &21.50 & $-0.95\pm0.51$ & $-5.63\pm0.5$  \\	         
7089 & M~2   & 21 33 27.3 & -00 49 23 & 11.5  & 10.4 &0.32 &12.75 & $+5.9\pm0.56$  & $-4.95\pm0.86$ \\
     & Pal~1 & 03 33 23.0 & +79 34 50 & 10.9  & 17.0 &2.28 &12.04 & \dots          & \dots          \\
     & Pal~3 & 10 05 31.4 & +00 04 17 & 92.7  & 95.9 &0.35 & 4.89 & $0.33\pm0.23$  & $0.30\pm0.31$  \\
     & Pal~4 & 11 29 16.8 & +28 58 25 & 109.2 &111.8 &0.33 & 3.30 & \dots          & \dots          \\
     & Pal~5 & 15 16 05.3 & -00 06 41 & 23.2  & 18.6 &2.28 &12.04 & $-1.78\pm0.17$ &$-2.32\pm0.23$  \\
     & Pal~14& 16 11 00.6 & +14 57 28 & 74.7  & 69.0 &0.77 & 7.20 & \dots          & \dots          \\
\hline
\end{tabular}
\end{table*}

\begin{figure*}
\centering
\includegraphics[height=6.5cm]{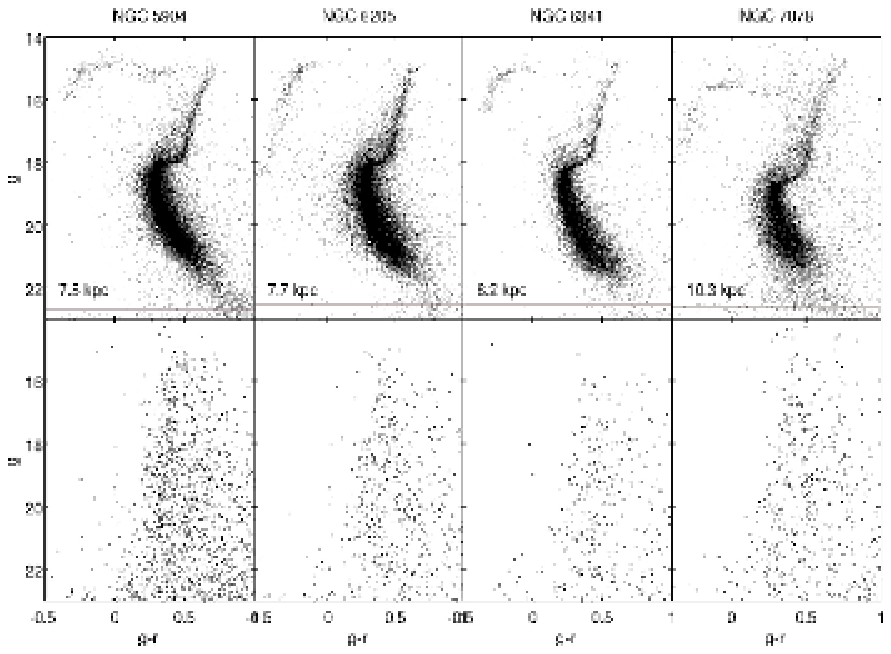}
\includegraphics[height=6.5cm]{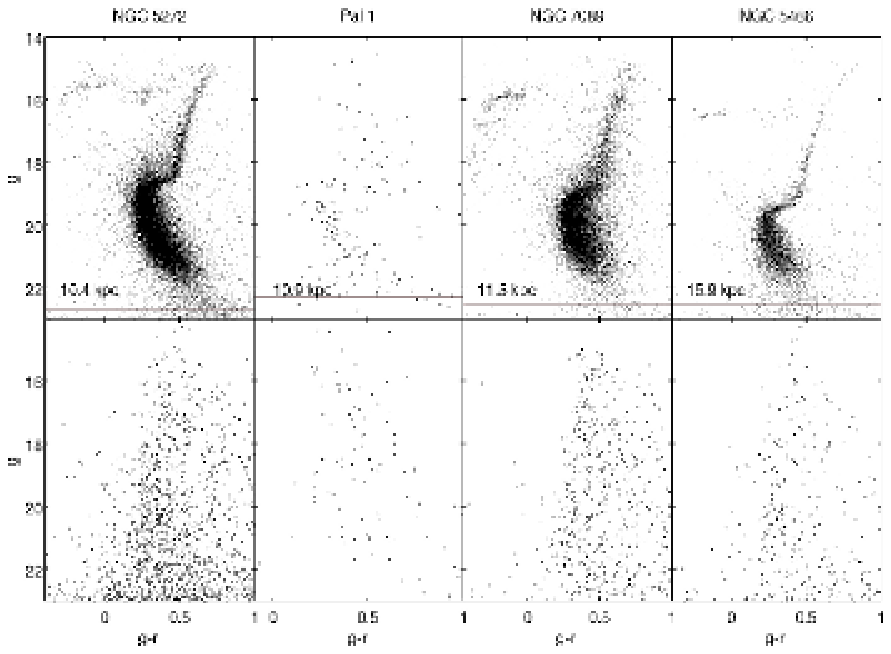}
\includegraphics[height=6.1cm]{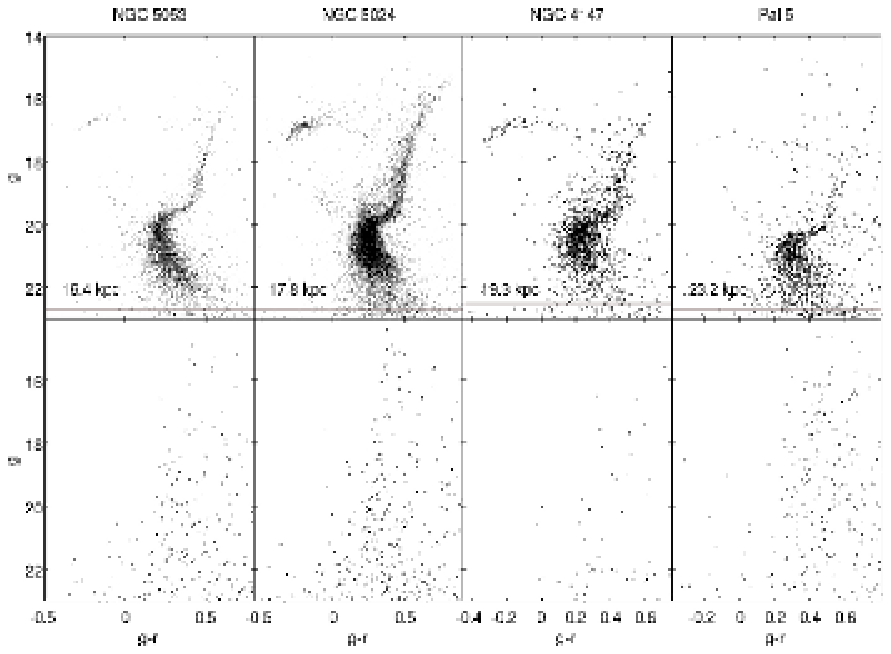}
\includegraphics[height=6.1cm]{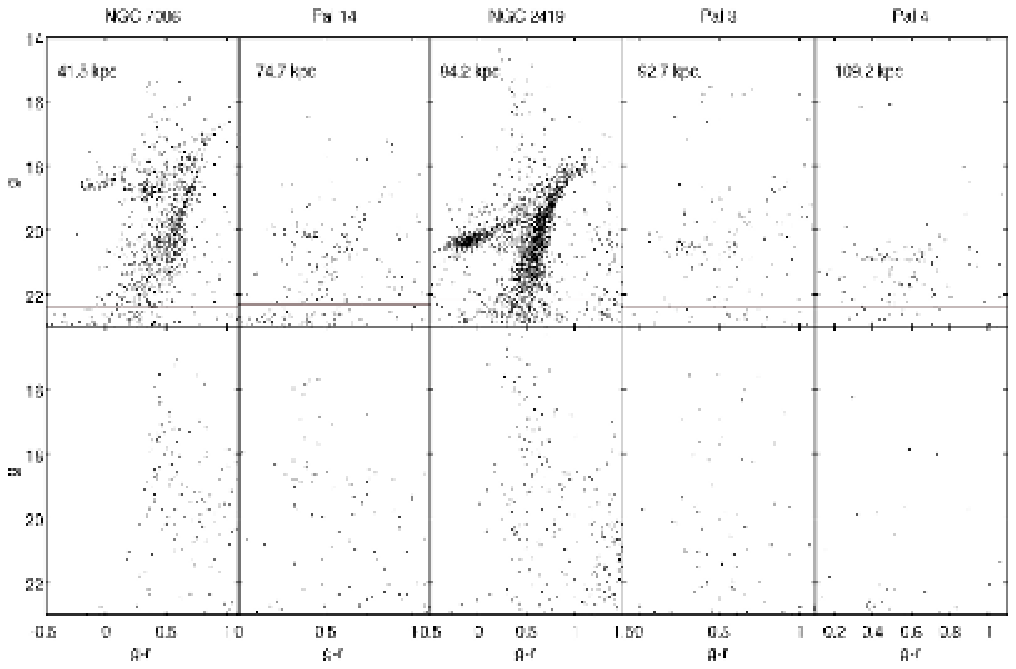}
\caption{\emph{Top rows:} Color-magnitude diagrams of all GCs in our sample in order of increasing distance from the Sun. The distances from \citet{harris96} are printed in each CMD. The grey horizontal line marks the magnitude limit for each cluster. \emph{Bottom rows:} Color-Magnitude diagrams of a random field in the vicinity of the cluster (see text for details).}
\label{Figcmdsall}
\end{figure*}

In Figure~\ref{Figcmdsall} we show the color-magnitude diagrams (g-r,g) of all GCs in our sample in order of increasing distance from the Sun (see column 5 in Table~\ref{tblgcs}). We show all stars within the clusters' tidal radii. In the bottom panels we show for each GC a sample of field stars. The field stars are chosen from a random field in the vicinity of the GC ($\sim 2^\circ$ away) within a circle equal to the clusters' tidal circle.

The SDSS data were taken over several years and in different observing conditions. The average seeing is $1.5\arcsec$ \citep{adelman07}. In our study we only use the g, r, and i band. \citet{ivezic04} showed that stellar population colors are constant over the survey area, concluding that the photometric calibration is accurate to roughly $0.02$~mag. Further, \citet{ivezic07} analyzed multiple observations of bright stars in the southern equatorial stripe. The photometry was repeatable to less than $0.01$~mag in all five bands. 

\section{Counting Algorithm}
\label{Seccounting}
To map cluster member stars on the sky we used the same method as \citet{odenkirchen03} to follow the tidal tails of Pal~5. We refer the reader to \citet{odenkirchen03} for a detailed description of the procedure. Here, we only give a rough overview. 

For the subsequent study of the distribution of stars on the sky we need a clean sample of potential member stars. In order to reduce the contamination of our sample with field stars we define new orthogonal color indices: 
\begin{eqnarray}
c_1&=&k_1\cdot(g-r)+k_2\cdot(g-i)+k_3\\
c_2&=&-k_2\cdot(g-r)+k_1\cdot(g-i)+k_4.
\end{eqnarray}

In a color-color plot the cluster stars are located in an almost one-dimensional locus. The new indices were now chosen in such a way that the color $c_1$ traces the cluster stars' locus and $c_2$ is perpendicular to that. To derive these indices for each GC we used the stars within $2/3\cdot r_t$, where $r_t$ is the cluster's tidal radius from Table~\ref{tblgcs}. We fitted a straight line \mbox{$y=\tan(\alpha)\cdot x+b$} to the distribution of stars in the color-color diagram $(g-r)~vs.~(g-i)$, where x represents the $(g-r)$ axis and y the $(g-i)$ axis. In Table~\ref{tblcolorindex} we list the fitted parameters $\tan(\alpha)$ and $b$ from which we derived $k_1=\cos(\alpha)$, $k_2=\sin(\alpha)$, $k_3=-b\cdot\sin(\alpha)$, and $k_4=-b\cdot\cos(\alpha)$ for each GC.

\begin{table*}
\caption{Parameters to derive the new color indices.}
\centering
\begin{tabular}{lcccccccccccc}
\hline
\hline
\noalign{\smallskip} 
NGC & 2419&4147&5024&5053&5272&5466&5904&6205&6341 \\
\hline
\noalign{\smallskip} 
$\tan(\alpha)$ &1.472&1.462&1.495&1.434&1.473&1.531&1.447&1.381&1.345 \\
$b$ & -0.039&0.004&-0.053&-0.006&-0.053&-0.047&-0.016&-0.010&-0.001\\  
\hline
\hline
\noalign{\smallskip} 
NGC/Name &7006&7078&7089  &Pal~1&Pal~3&Pal~4&Pal~5&Pal~14\\
\hline
\noalign{\smallskip} 
$\tan(\alpha)$ &1.540&1.522&1.473&1.545&1.404 &1.534&1.553&1.767 \\
$b$ &-0.028&-0.035&-0.054& 0.007&0.017&-0.147&-0.034&-0.190\\  
\noalign{\smallskip}     
\hline
\end{tabular}
\label{tblcolorindex}
\end{table*}

\begin{figure}
\centering
\includegraphics[width=9cm]{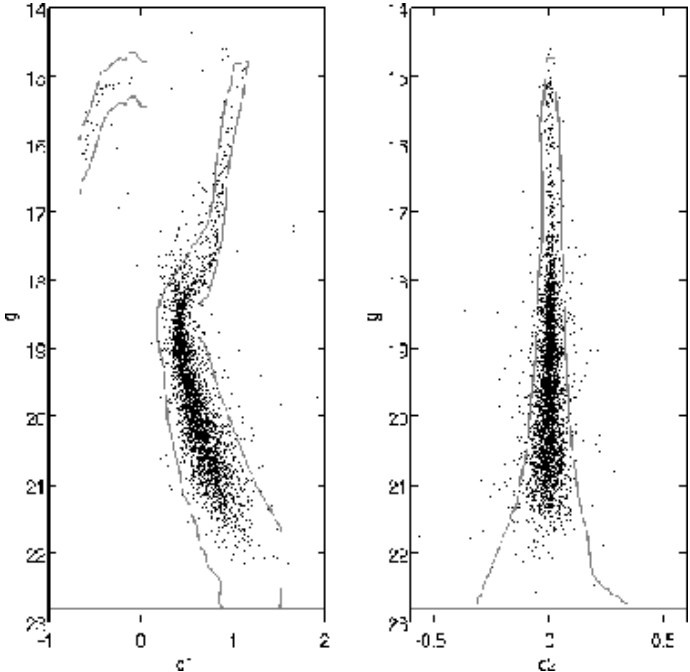}
\caption{Color magnitude diagrams (c$_1$,g) and (c$_2$,g). Plotted are stars of NGC~6341 within one third of the cluster's tidal radius. The gray solid lines denote the borders within which the stars were selected.}
\label{Figneuefarbe}
\end{figure}

We selected our stars in the color-magnitude diagrams ($c_1,g$) and ($c_2,g$) to minimize the contamination by Galactic field stars. Figure~\ref{Figneuefarbe} shows the CMDs ($c_1,g$) and ($c_2,g$) of NGC~6341. In ($c_2,g$) we selected all stars with $|c_2|<2.5\cdot\sigma(c_2)(g)$. $\sigma(c_2)(g)$ is the rms dispersion in $c_2$ for stars of magnitude g. Stars with deviating $c_2$ colors are unlikely to be cluster members.

In ($c_1,g$) we selected all stars within $3\sigma$ of the cluster's ridge line. The ridge line of a GC traces the location of highest density of stars in a color-magnitude diagram (CMD). Figure~\ref{Figneuefarbe} shows the selection criteria for NGC~6341 as an example. This preselection leaves us with the final catalog, where all stars are explicitly chosen to have the same photometric properties as the GC, but are located not only at the position of the cluster but also in the wide field around it. The final catalog is then split into the \emph{cluster sample} to contain all stars within $2/3\cdot r_{t}$ and the \emph{field sample} to contain all stars outside $3/2\cdot r_{t}$. If a cluster was observed to be more extended than previously thought, we redefined the two samples. In some cases, e.g., for NGC~5024 and NGC~5053, two clusters are close neighbors in the projection on the sky. In these cases, we excluded all stars within the tidal radius of the neighboring cluster from the \emph{field sample}.

\begin{figure}
\centering
\includegraphics[width=9cm]{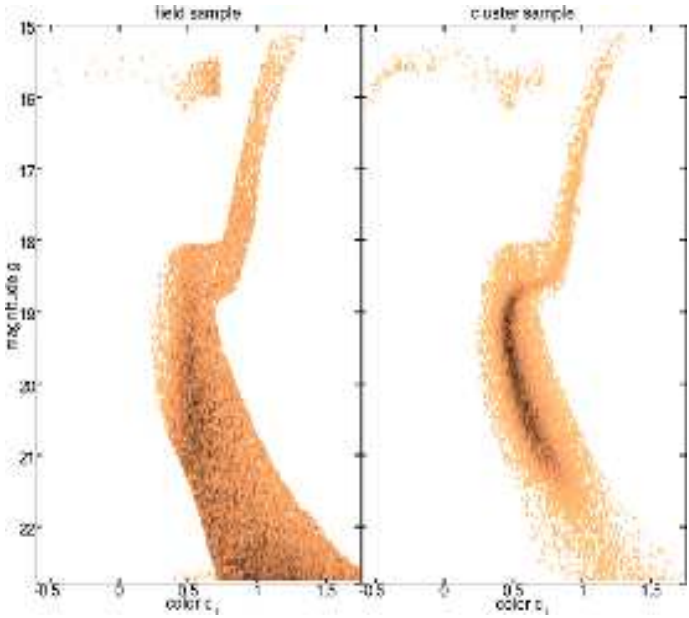}
\caption{Hess diagrams of the field (left panel) and cluster (right panel) sample of NGC~5272. Only stars already selected in c$_1$ and c$_2$ were used.}
\label{Figcmds}
\end{figure}

The stars in the preselected samples are not simply counted. The ratio of the distribution of cluster stars to field stars in the CMD ($c_1,g$) was used as a weight in the sum. Figure~\ref{Figcmds} shows the two Hess diagrams of NGC~5272 as contour plots. The left panel is the field sample, the right panel the cluster sample. The cluster sample shows a prominent main-sequence and subgiant branch. Furthermore, a red giant and horizontal branch are visible. For the field sample, the density is more homogeneous in the entire selected areas. The horizontal branch is only prominent at the red edge.

The density of cluster stars $n_C(k)$ at a given point $k$ on the sky is thus given as

\begin{eqnarray}
n_C(k)& =&  \frac{\sum_j n(j,k) \cdot \rho_C(j)/\rho_F(j)-n_F(k)}{\sum_j \rho_C^2(j)/\rho_F(j)},\label{eq}
\end{eqnarray}
where $\rho_C$ and $\rho_F$ are the density distribution of the cluster and field sample, respectively, in the CMD (see Figure~\ref{Figcmds}). $n_F$ is the average background in the field around the GC.

From Equation~\ref{eq} we see that each star in a given cell on the sky $k$ and in color-magnitude space $j$ is weighted according to its position in the CMD by the factor $\rho_C(j)/\rho_F(j)$. This weighted number of stars per cell on the sky is then summed and divided by the factor $a=\sum_j \rho_C^2(j)/\rho_F(j)$. This gives the estimated number of cluster stars $n_C$ plus a term $n_F/a$, the number of contaminating field stars attenuated by $a$. 

\section{Number density profiles}
\label{Secprofiles}
The large area coverage of the SDSS data allows us to study the number density profiles of GCs further out than any survey before. By using a large area ``far away'' from the cluster we are able to determine the mean background contamination and to investigate in detail the \emph{outer boundaries} of the GCs. 

We derived the number density profiles by counting the stars in concentric annuli of constant width $dr$. We used only the stars preselected in colors and magnitude as described above. We used annuli of width $2\arcmin$ outside the tidal radius. Within the cluster's tidal radius we chose either $0.5\arcmin$ or $1\arcmin$ as width of the annuli, depending on the surface density of the cluster. An alternative way to derive the number density profile would be by using annuli containing a constant number of stars. Comparing these two methods revealed no differences in the resulting profiles and parameters.

For the final number density profile we counted all stars in each annulus and divided by the area of the annulus. The error of the number density was derived by dividing the annulus in 20 segments. We then derived the number of stars in each segment and used the standard deviation of these 20 values as the error. For clusters with a small number of stars we divided the annuli in 8 segments to reduce the effect of small number statistics. 

For the denser clusters the SDSS is not suited to observe the number density profile in the cluster's central part due to crowding. Therefore, we combined the number density profiles from the SDSS with previously published profiles for the inner parts of the GCs. The surface brightness profiles by \citet{trager95} (T95) and the revised study of the same data by \citet{mclaughlin05} (M05) are the only large catalogs of surface brightness profiles of Galactic GCs. We combined these datasets with our new SDSS profiles to get profiles ranging from the cluster center out to several times the tidal radius. In order to combine the two datasets, we converted the surface brightness $\mu(r)$ of T95 into a number density $\log(N(r))=-\mu(r)/2.5+C$. $C$ is a constant derived separately for each GC in the following way. The T95 profile and the SDSS profile overlap within a certain radial range. This overlap was used to derive the vertical shift $C$ between the profiles. The T95 profiles were shifted by $C$ to match the SDSS profiles. For GCs whose T95 profiles extend far out, the outermost points were not taken into account, because these data points are the most affected by the background contamination or subtraction. Furthermore, differences in the outer parts of the different profiles can also be explained by mass segregation in the clusters, unveiled by different limiting magnitudes. 

For the GCs NGC~5272 (M3), NGC~5904 (M5), NGC~6205 (M13), NGC~6341 (M92), NGC~7078 (M15), and NGC~7089 (M2) \citet{noyola06} (N06) published surface brightness profiles derived from HST observations. They tested the influence of the chosen filter on the profiles and found no dependence. The profiles looked the same in each observed filter. For $50\%$ of their clusters the photometric data points are brighter in the central parts of the clusters than the fitted King profiles \citep{king62}. The discrepancy between the observations and the fit increases as the radius decreases. N06 conclude that the inner parts are not suited to be fitted with a King profile. 

For these six GCs we combined the N06 profiles and our SDSS profiles to span the largest possible range in radius. The combination was done in an identical way as for the combination of the T95 sample with our SDSS profiles.

\begin{figure*}[tbph]
\centering
\includegraphics[width=\textwidth]{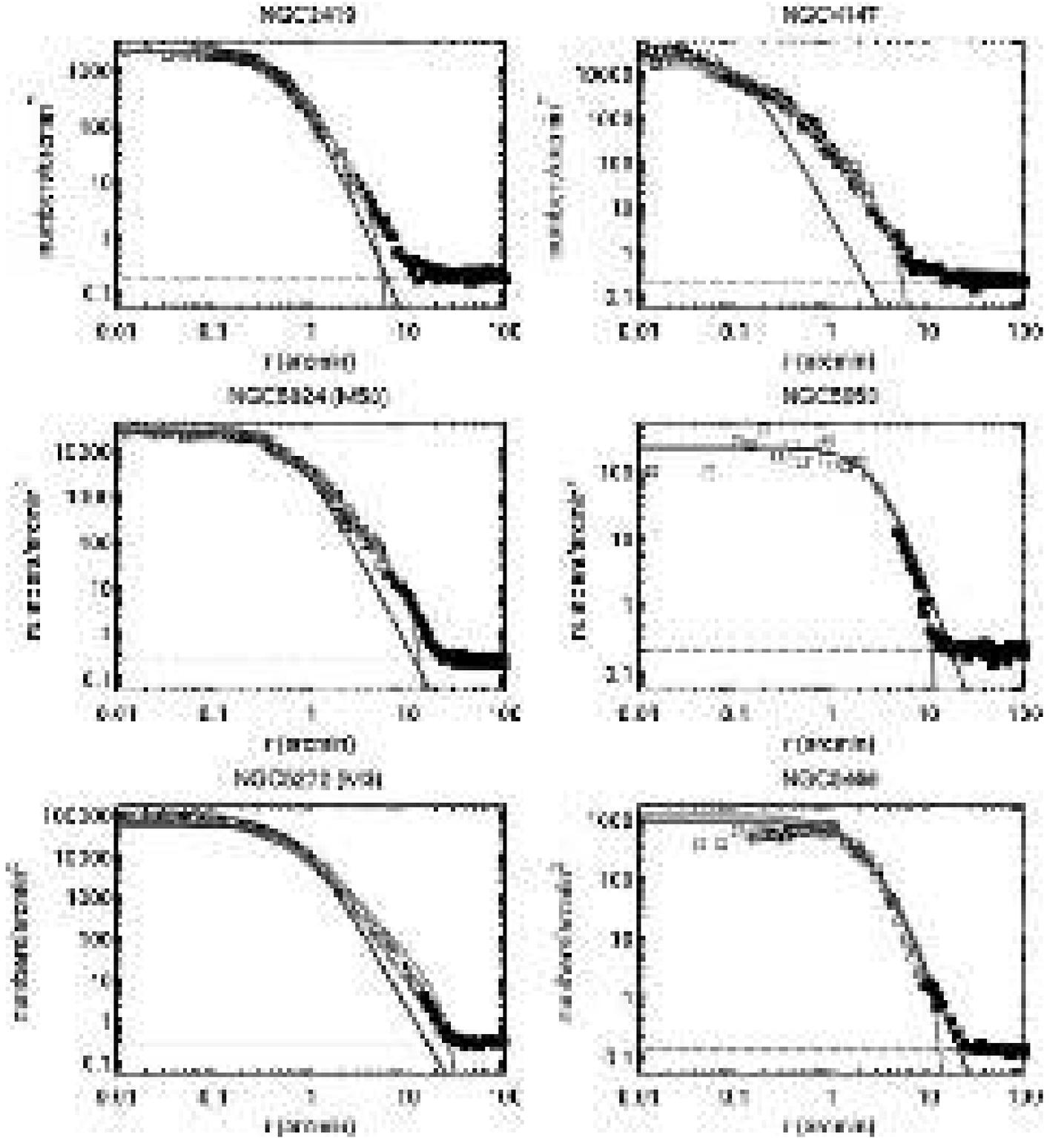}
\caption{Number density profiles of NGC~2419, NGC~4147, NGC~5024, NGC~5053, NGC~5272, and NGC~5466. The black open squares show the combined profiles. The black filled squares were derived in this study. The light grey solid line is the best fit King profile. The dark grey line is the best fit Plummer profile. The horizontal line marks the measured, mean background $n_{bkg}$.}
\label{Figprofiles1}
\end{figure*}

\begin{figure*}[tbph]
\centering
\includegraphics[width=\textwidth]{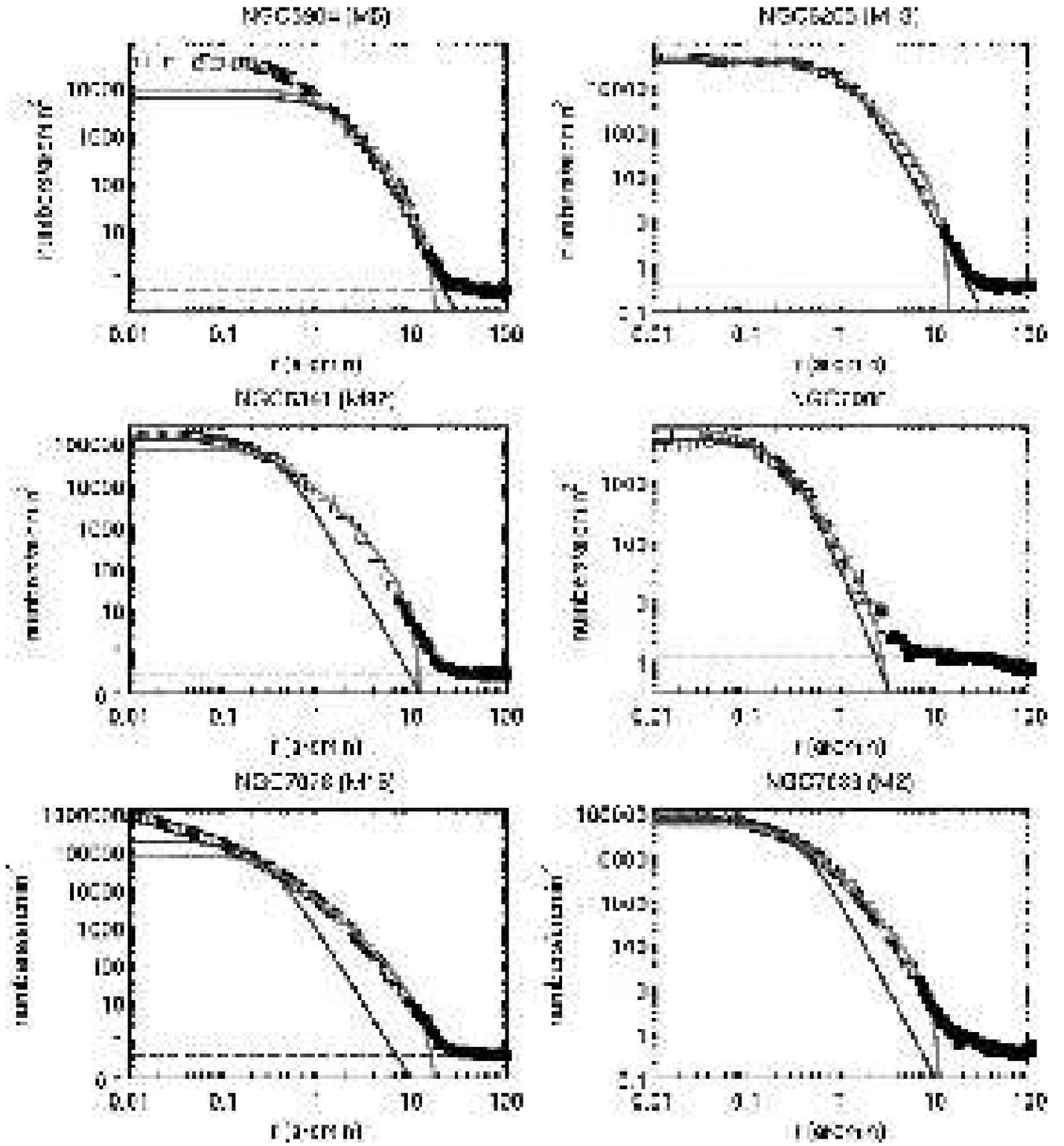}
\caption{Number density profiles of NGC~5904, NGC~6205, NGC~6341, NGC~7006, NGC~7078, and NGC~7089. The black open squares show the combined profiles. The black filled squares were derived in this study. The light grey solid line is the best fit King profile. The dark grey line is the best fit Plummer profile. The horizontal line marks the measured, mean background $n_{bkg}$ .}
\label{Figprofiles2}
\end{figure*}

\begin{figure*}[tbph]
\centering
\includegraphics[width=\textwidth]{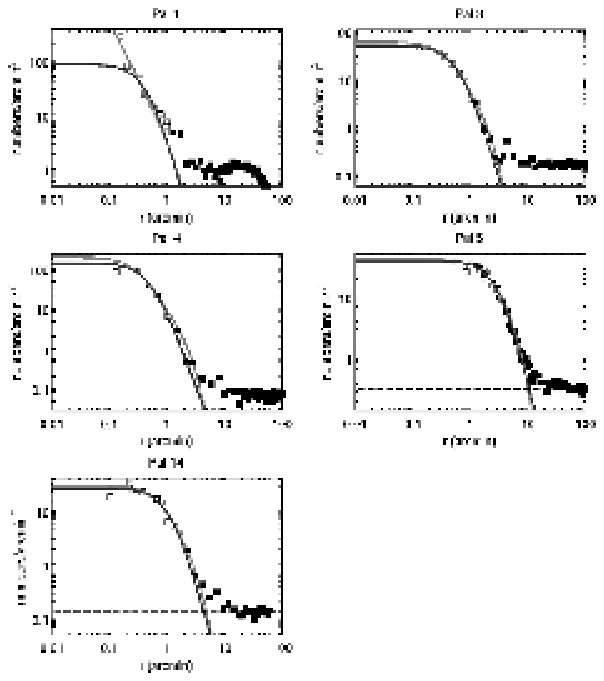}
\caption{Number density profiles of Pal~1, Pal~3, Pal~4, Pal~5, and Pal~14. The black open squares show the combined profiles. The black filled squares were derived in this study.  The light grey solid line is the best fit King profile. The dark grey line is the best fit Plummer profile. The horizontal line marks the measured, mean background $n_{bkg}$.}
\label{Figprofiles3}
\end{figure*}

The combined profiles were fitted with \citet{king62} and \citet{plummer11} profiles. For the King profiles, we adopted the background levels derived from the SDSS observations and determined for the core and tidal radius through our fit. For the Plummer profiles we derived the Plummer radius $r_p$ through our fit. 
In Figures~\ref{Figprofiles1}--\ref{Figprofiles3} we show the combined number density profiles of all clusters in our sample. Table~\ref{tabparam} lists the fitted core and tidal radii, the observed stellar background density $n_{bkg}$ and the Plummer radius. For some GCs, e.g. NGC5466, NGC~6205, the Plummer profile provides a better fit than the King profile. But in most cases, the Plummer profile does not trace the observed number density profile. \citet{penarrubia09} found that a Plummer profile more accurately traces the profile of a stellar system (mainly a dwarf spheroidal galaxy) that loses mass due to tidal effects. For NGC~5466 this is certainly the case and the known tidal tails are proof of tidal mass loss. For Pal~5, which is also known for its long tidal tails, the Plummer profile is not a better fit than the King profile. In the following Section we will discuss the prominent features and trends in the density profiles for each cluster individually.

\begin{table*}
\centering
\caption{Measured structural parameters.\label{tabparam}}
\begin{tabular}{llcccc}
\hline
\hline
\noalign{\smallskip}  
{NGC}&Name &$n_{bkg}$&  {$r_c$}  & {$r_t$} & {$r_p$}\\
 & &stars/arcmin$^2$& arcmin & arcmin & arcmin\\
\noalign{\smallskip}     
\hline
2419 &       &0.35& $0.30\pm0.05$ &$ 8.07\pm0.87$&$0.59\pm0.01$  \\
4147 &       &0.20& $0.14\pm0.01$ &$ 6.26\pm1.53$&$0.12\pm0.01$  \\
5024 & M~53  &0.22& $0.37\pm0.01$ &$14.79\pm7.19$&$0.66\pm0.01$  \\
5053 &       &0.08& $2.23\pm0.36$ &$12.52\pm4.49$&$3.00\pm0.3$  \\
5272 & M~3   &0.17& $0.41\pm0.003$&$31.81\pm12.3$&$0.77\pm0.01$  \\
5466 &       &0.03& $1.19\pm0.09$ &$14.63\pm2.40$&$2.32\pm0.08$  \\
5904 & M~5   &0.52& $1.04\pm0.06$ &$18.91\pm2.71$&$2.18\pm0.05$  \\
6205 & M~13  &0.33& $0.82\pm0.005$&$14.36\pm0.96$&$1.27\pm0.01$  \\
6341 & M~92  &0.15& $0.37\pm0.01$ &$12.55\pm0.90$&$0.36\pm0.01$  \\
7006 &       &1.14& $0.14\pm0.01$ &$ 3.64\pm0.56$&$0.29\pm0.01$  \\
7078 & M~15  &0.16& $0.27\pm0.03$ &$18.35\pm1.97$&$0.26\pm0.01$ \\
7089 & M~2   &0.13& $0.29\pm0.002$&$11.72\pm0.63$&$0.35\pm0.01$  \\
     & Pal~1 &0.13& $0.005\pm1.06$&$ 4.09\pm0.92$&$0.49\pm0.03$ \\
     & Pal~3 &0.03& $0.37\pm0.02$ &$ 5.75\pm0.92$&$0.65\pm0.01$ \\
     & Pal~4 &0.15& $0.26\pm0.10$ &$ 5.30\pm0.65$&$0.54\pm0.01$ \\
     & Pal~5 &0.15& $2.34\pm0.14$ &$16.03\pm2.24$&$3.38\pm0.11$  \\
     & Pal~14&0.07& $0.72\pm0.08$ &$ 8.82\pm3.20$&$1.19\pm0.06$ \\
\hline
\end{tabular}
\end{table*}

In Figures~\ref{Fig2419}--\ref{FigPal14} we plot our newly derived tidal radii as red ellipses. The tidal radius, $r_t$, from the \citet{king62} profiles is the radius at which the stellar density (number of stars per arcmin$^2$) drops to zero. It must not necessarily be the radius outside of which a star has left the cluster potential, but it is an estimate of this radius. The radius outside of which the star left the cluster potential is the distance between the cluster center and the first Lagrangian point, the Jacobi radius \citep{baumgardt09}. Stars outside the Jacobi radius have left the GC's potential. \citet{baumgardt09} calculated Jacobi radii for all Galactic GCs. We adopted these values. In Figures~\ref{Fig2419}--\ref{FigPal14} the Jacobi radii are plotted as blue dashed ellipses.

\section{Tidal Features}
\label{Sectidal}
In all contour maps we plot iso-density lines at levels of 
1,2,3,5,7,8,10,20,40,60,80,100,200,400,600,800,1000,2000$\cdot \sigma_{bkg}$ above the mean weighted background level $n^w_{bkg}$. The mean weighted background was derived at least two tidal radii away from the cluster and if there were unusable areas, e.g., close neighboring clusters, overlapping SDSS scans, areas outside the SDSS footprint, known tidal tails, etc., these were not taken into account for its determination. $\sigma_{bkg}$ denotes the standard deviation of the weighted density in the area, which was used to determine $n^w_{bkg}$. The mean weighted background, $n^w_{bkg}$, is different from the background $n_{bkg}$ derived in Section~\ref{Secprofiles}. $n_{bkg}$ is the mean of the number of stars per area, whereas for $n^w_{bkg}$ the stars were weighted by the factor $\rho_C(j)/\rho_F(j)$ according to their position in the CMD. 

In Figures~\ref{Fig2419}--\ref{FigPal14} we present the contour maps for all 17 GCs in our sample. The upper/left panel shows a large area, typically $9\deg$ by $9\deg$, in order to show potential large scale features in a Lambert conical projection. The lower/right panels show a blow-up in order to study the small scale tidal structures of the GCs. On the x-axis of the zoom-in plots we plot the coordinate $RA_{shown}=X_m+(RA-X_m)\cdot \cos(Y_m)$, where ($X_m$,$Y_m$) is the center of the cluster. All contour plots have North up and East right. The calculated iso-density contours are smoothed with a Gaussian kernel over $5$~pixels ($=15$~arcmin) with a sigma of $2$~pixels ($=6$~arcmin). In all maps the $1\sigma_{bkg}$ and the $2\sigma_{bkg}$ are shown in grey, while the higher contours are shown in black.

Before presenting all contour maps, we briefly discuss the possible sources for contamination. 

\begin{itemize}
\item{
\textbf{Background galaxies \& quasars}\\
The SDSS automatic Star/Galaxy separation seems to be very reliable. As we only use data considered to be a point source we, in principle, do not have to worry about contaminating background galaxies or galaxy clusters. In the fields with An08 photometry this may be an issue, but these fields do have a high stellar density, with stars outnumbering contaminating galaxies.

As quasars are point sources and are included in the SDSS database together with stars, these might influence our data. We have looked at the SDSS quasar catalog IV \citep{schneider07}, which contains $77\,429$ quasars from SDSS DR5. We applied the same selection criteria in color and magnitude to select possible contaminating quasars for each GCs, respectively. E.g., for NGC~5272 we end up with $54$ quasars contaminating our sample of $51\,964$ stars in total. The contamination due to quasars can be neglected.}

\item{
\textbf{Dust}\\
Extinction by dust might also influence our tidal features. All our magnitudes are corrected for extinction as given in the SDSS database by values from \citet{schlegel98}. In a first look, we do not expect a strong contamination by dust, as the SDSS mainly observed the sky around the North Galactic pole (NGP), where less dust is expected. But, in previous studies a correlation between tidal tails and dust was found for some GCs. We examined for each GC the orientation of the tidal features and studied the potential correlation with the dust. We did not observe any correlation.}

\item{
\textbf{Foreground \& background stars}\\
In Section~\ref{Seccounting} we explained our criteria to select stars in color-magnitude space to minimize the number of stellar foreground and background contaminants. For all GCs we have only selected stars within $3\sigma$ of the cluster's ridge line to assure that the majority of stars counted are likely cluster members. The 2d-distribution of these selected stars shows in all cases a unique maximum at the GC's position. The area around the GC is in all cases not perfectly flat, but shows density peaks of various significances. I.e., we are able to clean our stellar sample, but not perfectly. A more accurate exclusion of Galactic foreground and background contaminants can only be achieved with spectroscopic confirmations for each single star - a task which seems almost impossible for the time being.}
\end{itemize}

\subsection{NGC~2419}
NGC~2419 is a very bright, unusually large GC located at a very large distance from the center of the Milky Way ($R_{MW}=91.5$~kpc, H96). Previous studies of its surface brightness profile revealed some evidence for extratidal stars \citep{bellazzini07} and references therein). In Figure~\ref{Figprofiles1} our combined number density profile is shown. We also observe a break in the profile's slope. No proper motion information is available. As NGC~2419 is very remote, only the upper part of the red-giant branch and the horizontal branch are visible in the SDSS data. 

\begin{figure*}
\centering
\includegraphics[height=8.5cm]{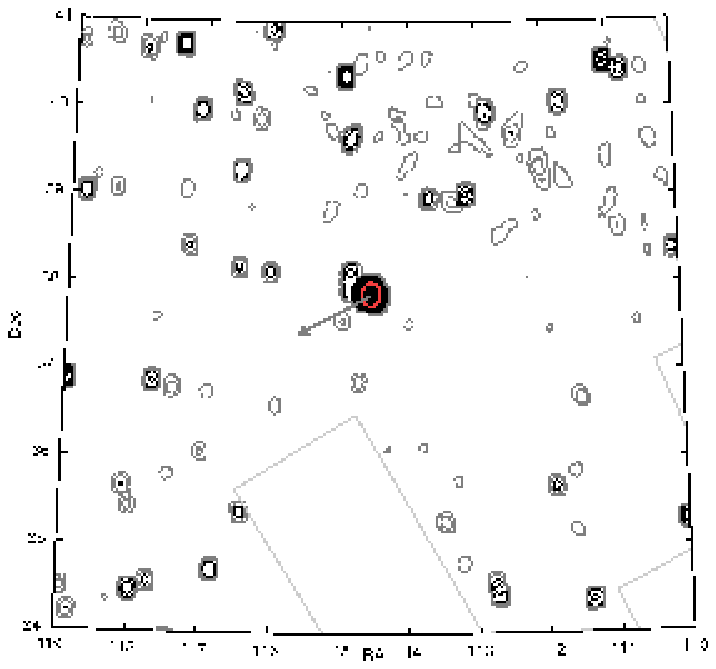}
\includegraphics[height=8.5cm]{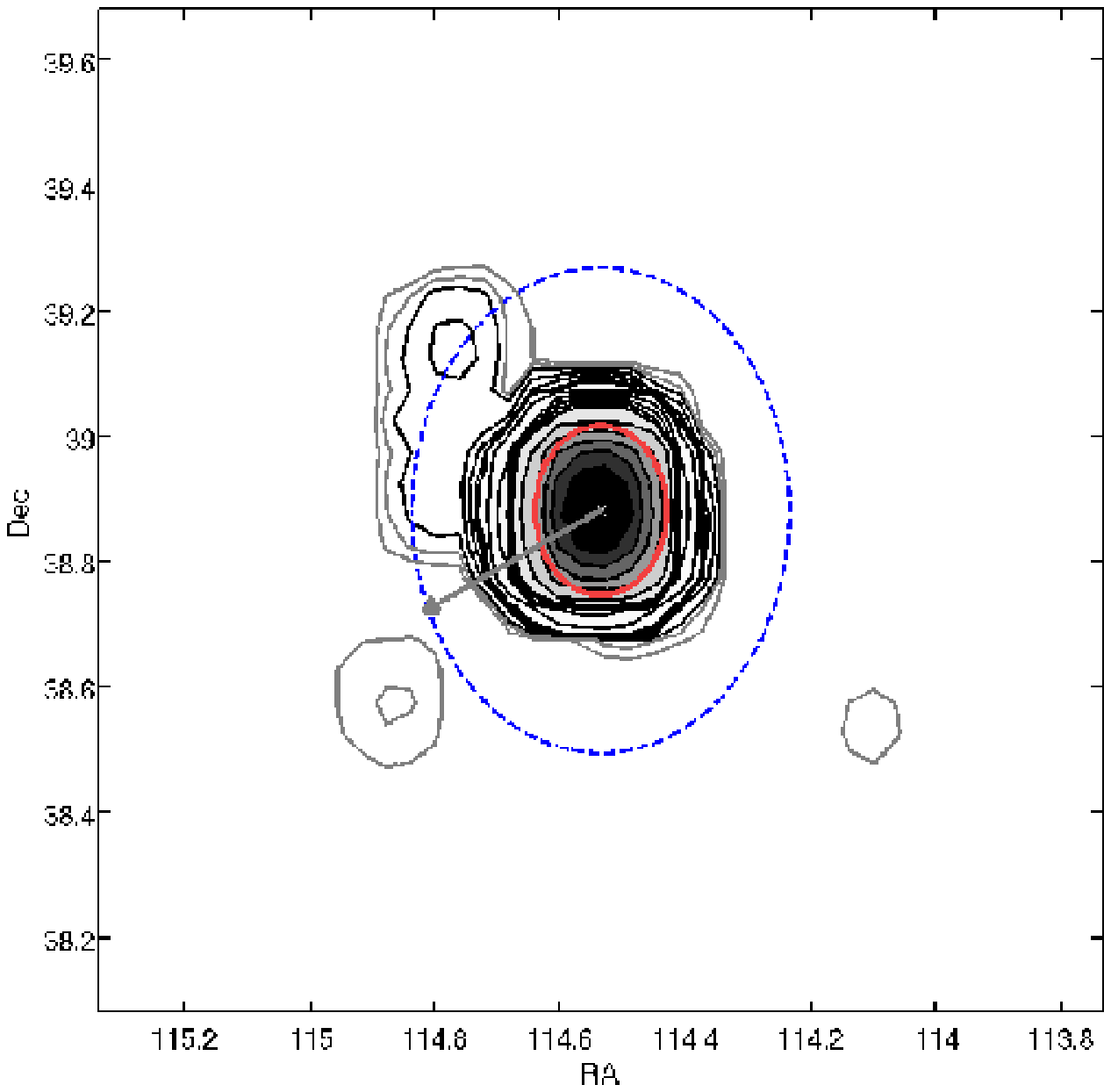}
\caption{\emph{Left panel:} Contour plot of a large area centered on NGC~2419. \emph{Right panel:} Blow-up of the contour plot centered on NGC~2419. In both panels the isodensity lines are drawn at levels of 1, 2, 3, 5 ,7 ,8 ,10, 20, 40,$\dots \cdot \sigma_{bkg}$ above the mean weighted background level. The $1\sigma_{bkg}$ and $2\sigma_{bkg}$ contours are drawn in light grey, the higher levels are drawn in black. The solid grey arrow points towards the Galactic Center. The grey solid line marks the edge of the SDSS survey area. The red solid circle shows the King tidal radius. The blue dashed circle shows the Jacobi radius.}
\label{Fig2419}
\end{figure*}

A stripe north of NGC~2419 has an artificially doubled background density. We corrected for this. All stars in the area denoted by the dashed grey line were studied. If two stars were closer than $2\arcsec$ and similar in color and magnitude one of the stars was deleted from the sample. In Figure~\ref{Fig2419} we show the resulting contour map. The average background level is $1.21\cdot10^{-1}$~stars~arcmin$^{-2}$. We detect overdensities several $\sigma$ above the mean background. These features show sharp edges and a flat central plateau. The GC has a pronounced extratidal halo, but is mostly contained within its Jacobi radius. In the northeast we observed distorted contours. Comparable  features might be visible on the cluster's southwestern side, but on larger scales no features connected to NGC~2419 are observed. \citet{casetti09} speculated that NGC~2419 might be the nucleus of a (former) dwarf galaxy with the Virgo Stellar Stream being its tidal tail.

\subsection{NGC~4147}
\begin{figure*}
\centering
\includegraphics[height=8.5cm]{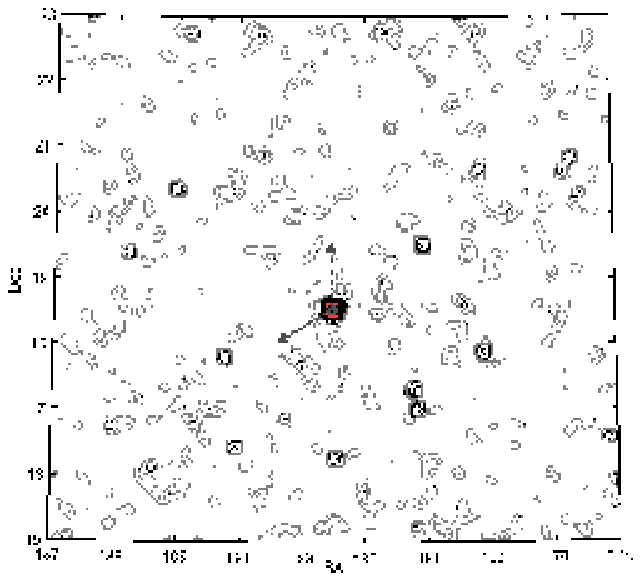}
\includegraphics[height=8.5cm]{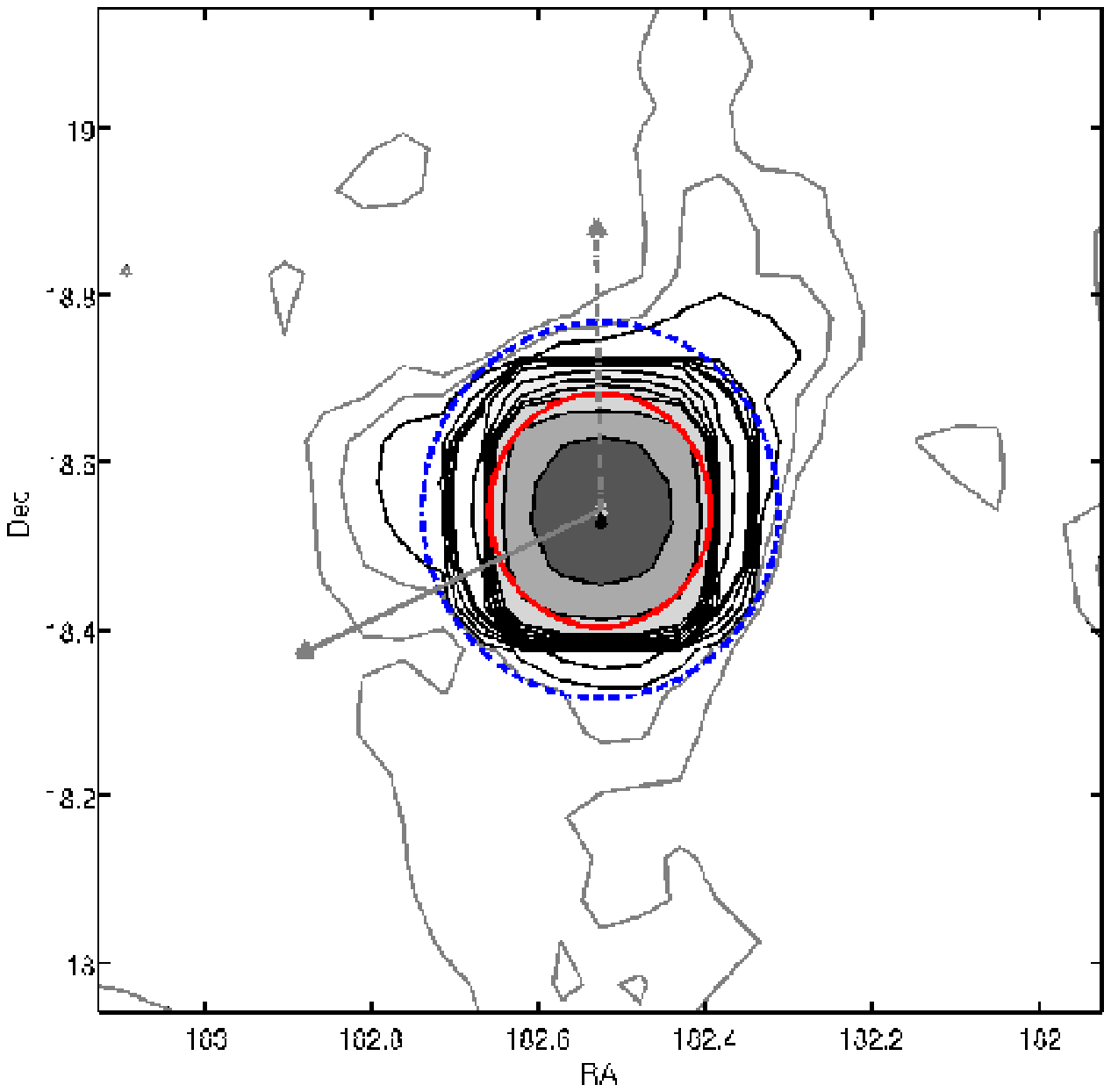}
\caption{Same as Figure~\ref{Fig2419} but for NGC~4147. The grey dashed arrow indicates the direction of its Galactic motion based on the proper motions stated in Table~\ref{tblgcs}.}
\label{Fig4147}
\end{figure*}

From the position in the vital diagram (in Figure~\ref{figVital}) we expect to observe some hint for the cluster's dissolution due to relaxation. NGC~4147 is currently about $18.7$~kpc from the Galactic center, close to apogalacticon $R_{apo}=25.5$~kpc (D99). In Figure~\ref{Fig4147} we plot the contour map of the cluster with a weighted background level of $n^w_{bkg}=3.89\cdot10^{-1}$~stars~arcmin$^{-2}$. The large area plot reveals a variable background. Overdensities of several $\sigma$ above $n^w_{bkg}$ are detected. The close-up reveals a complex multiple arm morphology. The $1\sigma$ contour appears to indicate $\mathrm{S}$-shaped tidal arms extending over several tidal radii in southern and northern direction. We detect a halo of extratidal stars and for the three lowest contours we observe multiple tidal arms. Such a multiple arm morphology is predicted by \citet{montuori07} for clusters on elliptical orbits ($e_{4147}=0.73$) close to apogalacticon. NGC~288 \citep{leon00} and Willman I \citep{willman06} have a similarly complex morphology, showing three tidal arms. \citet{martinez04} investigated the CMD of NGC~4147 and of a comparison field close by. They detected some hints for extratidal members stars. In the upper right plot of Figure~\ref{Figprofiles1} we plotted the number density profile of NGC~4147. A break in the profile's slope is observed. 

\citet{bellazzini03} found stars of the Sagittarius tidal arm in the vicinity of NGC~4147. They conclude that this cluster might be connected to the Sagittarius dwarf (Sgr dwarf). Recently, \citet{law10} investigated the association of MW GCs with the Sgr dwarf based on the authors' newest Sgr model. The authors conclude, in contrast to earlier studies \citep[e.g.,][]{bellazzini03,forbes10}, that NGC~4147's association with the Sgr dwarf is of relative low confidence. It is thus unclear whether the extended features we observe are a feature of the Sgr dwarf field population. In any case, these stars are located at a similar distance as the cluster.

\subsection{NGC~5024 (M~53)}
\begin{figure*}
\centering
\includegraphics[height=9cm]{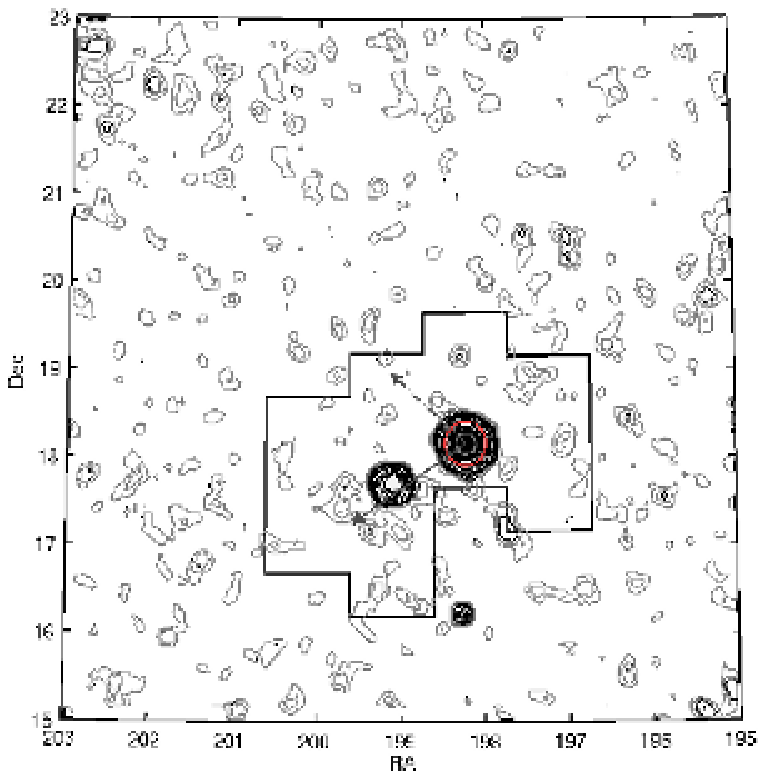}
\includegraphics[height=9cm]{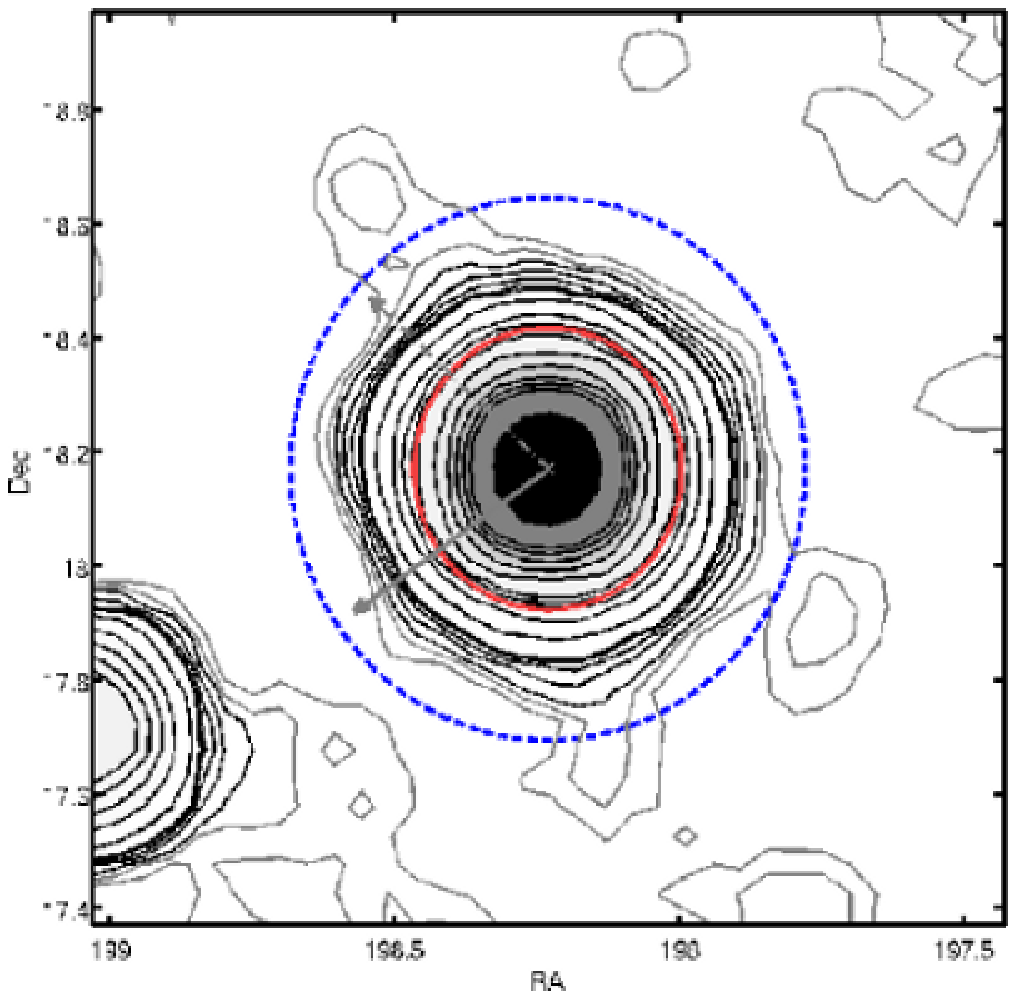}
\caption{Same as Figure~\ref{Fig2419} but for NGC~5024. The large overdensity in the left panel at (199,17.8) is the GC NGC~5053. The solid black line denotes the area observed by \citet{chun10}.}
\label{Fig5024}
\end{figure*}

NGC~5024 is, in projection on the sky, a very close neighbor of NGC~5053, although the two clusters are more than $1$~kpc away from each other in radial direction. NGC~5024 is currently located about $18.4$~kpc from the Galactic center close to its perigalacticon $R_{peri}=15.7$~kpc (D99). In the middle left plot of Figure~\ref{Figprofiles1} its number density profile is shown. The King model fitted to the combined profile reveals extratidal stars, a mild break in the slope is visible. In Figure~\ref{Fig5024} we plot the contour map of the cluster. In the left panel, the contour map of the entire observed area is shown. The cluster is clearly detected as the highest density peak. In the surrounding field mostly features up to $3\sigma$ above the mean of $0.6$~stars~arcmin$^{-2}$ are observed. Southeast of NGC~5024 the close neighbor NGC~5053 is detected, but no connection between the two is observed in contrast to the findings of \citet{chun10}. The black solid polygon shows the area observed by \citet{chun10}. We find a strong ($9\sigma$) overdensity at $(198.2,16.3)$. In Figure~\ref{Fig5053} this overdensity is not recognizable anymore. Therefore, we argue that it is an overdensity at the distance of NGC~5024. We found exactly one possible BHB/RR~Lyrae star centered on the overdensity. It is still debated in the literature whether NGC~5024 is associated with the Sgr tidal stream \citep[see][]{bellazzini03b,law10}. Our data do not permit us to conclude whether the overdensity is related to NGC~5024 or the Sgr stream. We found no obvious large scale structure connected with NGC~5024. The zoom-in on NGC~5024 shows interesting features. The contours are spherical and smooth not only in the cluster center, but also in the pronounced extratidal halo. The $1\sigma$ contour is strongly influenced by the background. \citet{chun10} also observed NGC~5024 and NGC~5053. They detected a tidal-bridge like feature around the two GCs and tidal features that may be due to dynamical interaction between the two clusters. Their findings are in contrast to our observations.

\subsection{NGC~5053}
\begin{figure*}
\centering
\includegraphics[height=8.5cm]{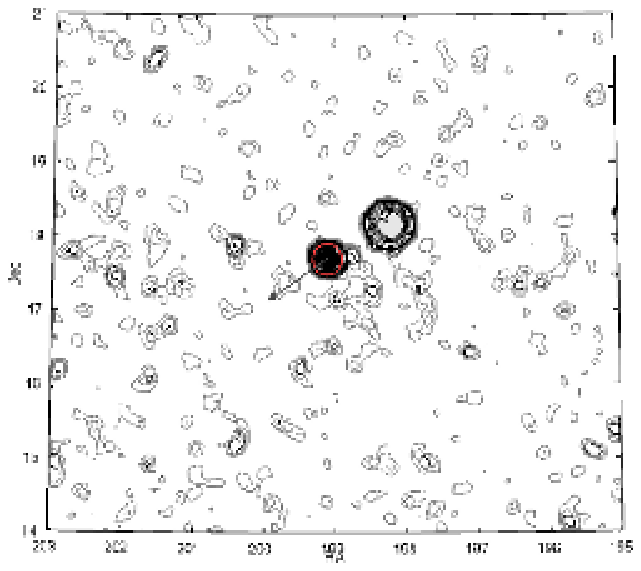}
\includegraphics[height=8.5cm]{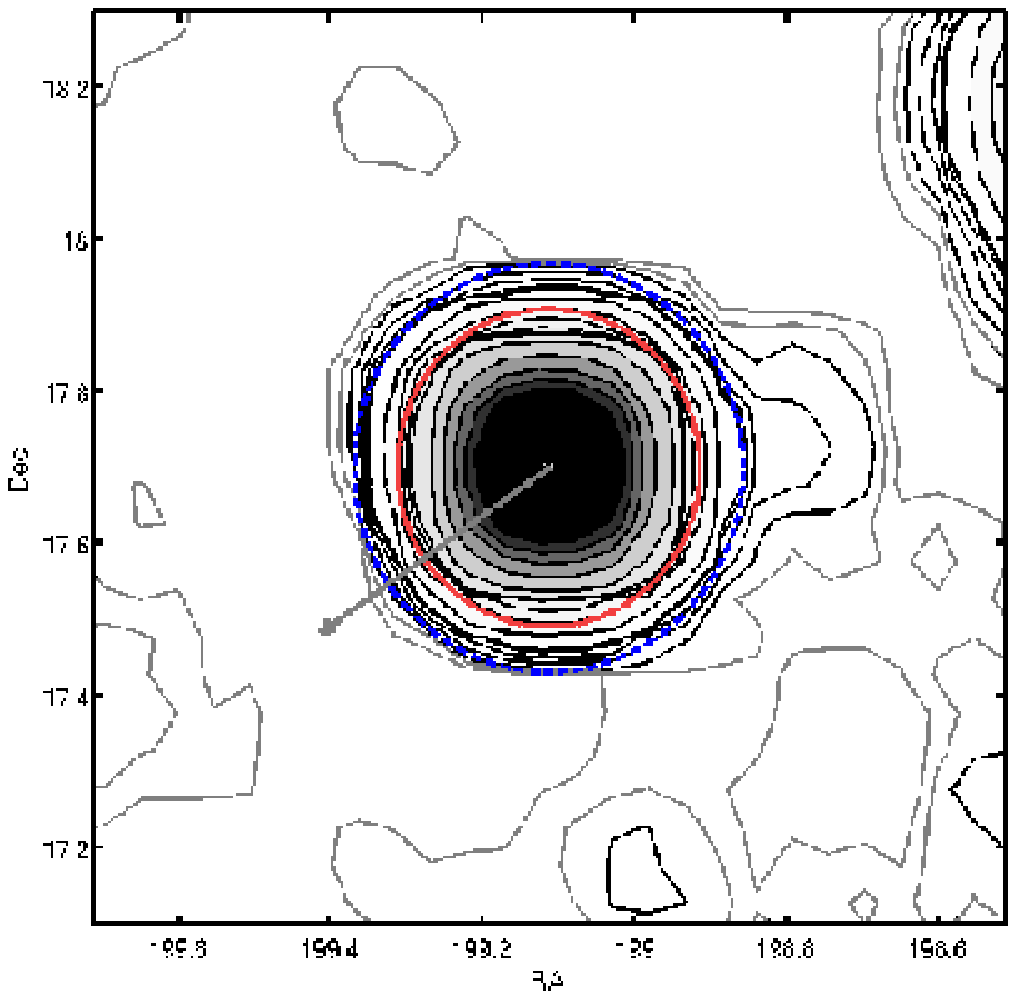}
\caption{Same as Figure~\ref{Fig2419} but for NGC~5053. The large overdensity in the left panel at (198,18) is the GC NGC~5024.}
\label{Fig5053}
\end{figure*}

NGC~5053 located about $16.9$~kpc from the Galactic center. \citet{lauchner06} studied the 2d distribution of NGC~5053 and detected extratidal features east and west of the cluster. NGC~5053 is believed to be a member of the Sgr dSph \citep{bellazzini03b}. In Figure~\ref{Fig5053} we show its contour map. In the large area contour map we see no pronounced large scale structure, but we definitely detect the GC and also random background noise up to $3\sigma$ above the mean of $4.1\cdot10^{-2}$~stars arcmin$^{-2}$. The large overdensity northeast of the cluster is NGC~5024. The right panel of Figure~\ref{Fig5053} is a zoom-in on NGC~5053. It reveals distorted contour lines. We observe a halo of extratidal stars. Towards the northeast and west the contours show a strong asymmetry. Similar features were observed by \citet{lauchner06}. We see a one-armed tail-like extension toward the north-west. Such a structure is also seen to emerge from NGC~5024 (both in Figures~\ref{Fig5024} \& \ref{Fig5053}). Our data do not permit us to tell whether these curious one-armed features are tidal tails.  \citet{law10} classify NGC~5053 as moderately likely associated with the Sgr dwarf. Therefore, it is possible that we are looking at substructure in the Sgr dwarf field population. \citet{forbes10} discuss the possibility that NGC~5053 or NGC~5024 might be the remnant nucleus of a dwarf galaxy. The number density profile is shown in the middle, right panel of Figure~\ref{Figprofiles1}. It is interesting to note that the density profile (Figure~\ref{Figprofiles1}) does not reveal a pronounced extratidal feature. Also \citet{lehmann97}, on photographic plates, did not find any indications for tidal tails. \citet{chun10} observed NGC~5053 together with NGC~5024, and as stated earlier we do not reproduce their findings.

\subsection{NGC~5272 (M~3)}
\begin{figure*}
\centering
\includegraphics[height=8.5cm]{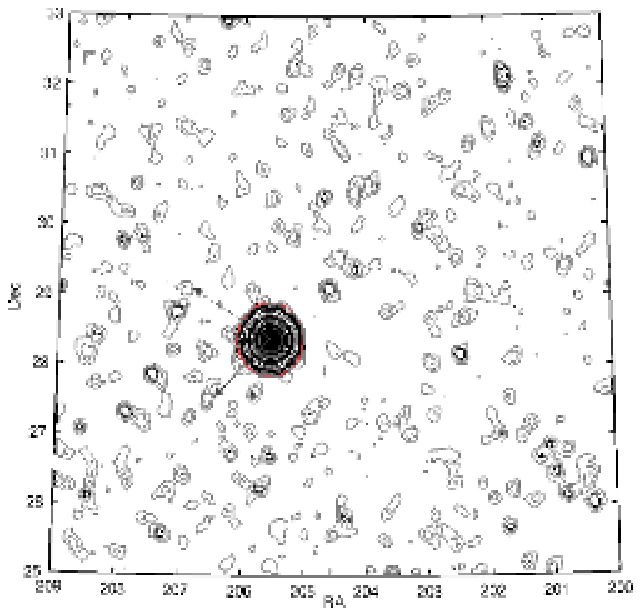}
\includegraphics[height=8.5cm]{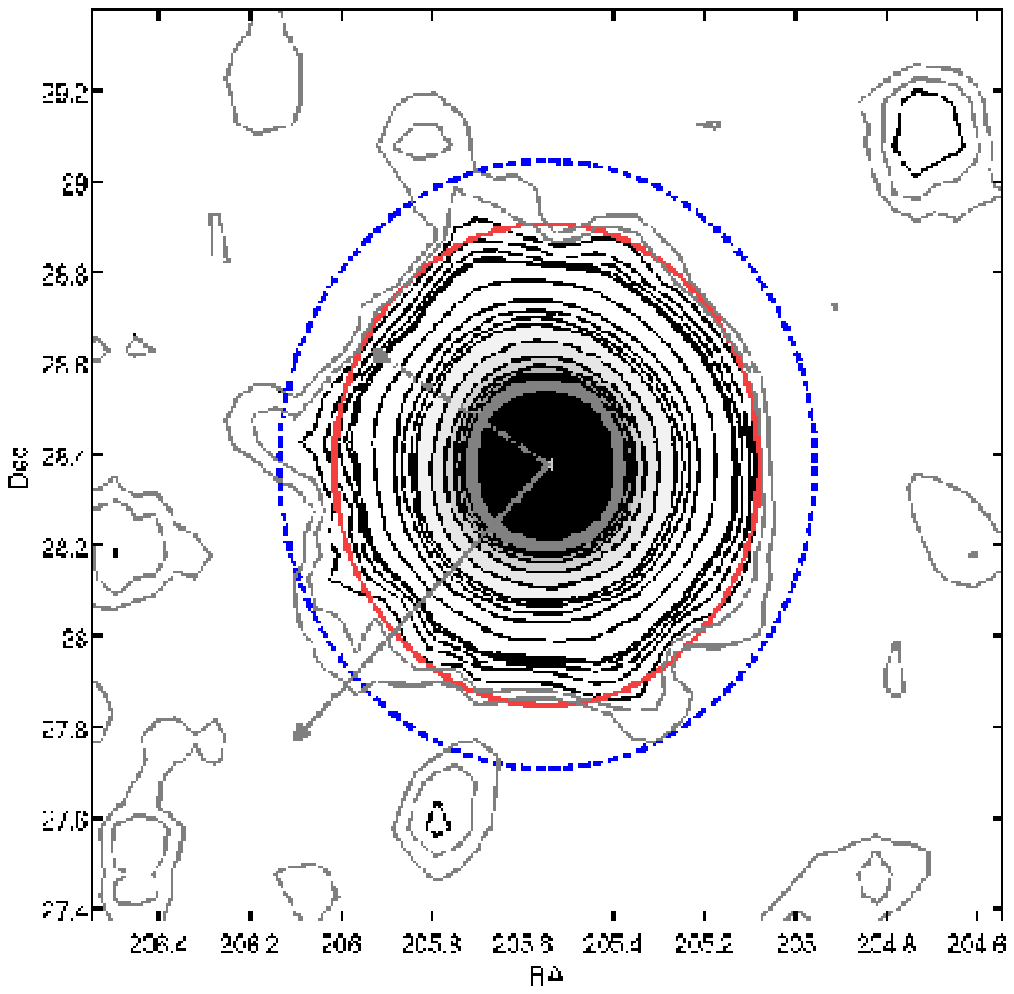}
\caption{Same as Figure~\ref{Fig2419} but for NGC~5272.}
\label{Fig5272}
\end{figure*}

NGC~5272 is $11.5$~kpc away from the Galactic center, located close to apogalacticon $R_{apo}=13.7$~kpc. Compared to other clusters in our sample NGC~5272 has a low expected destruction rate (GO97). The lower left panel of Figure~\ref{Figprofiles1} shows its number density profile. The combined profile is not well fit with a King profile. The profile shows no change in slope as a hint for an extratidal halo. In Figure~\ref{Fig5272} we show the contour map of this cluster. Generally, the contours are smooth. On the other hand the outer contours up to $3\sigma$ show distortions. The extreme distortions of the lowest contour are likely to be induced by the background. There is no large scale tidal structure observed. The background shows random density peaks of up to $3\sigma$ above the mean of $4.45\cdot10^{-2}$~stars arcmin$^{-2}$. \citet{leon00} studied the 2d distribution of NGC~5272 and found extratidal features correlated \textbf{with} dust emission. Our contours do not show any correlation with the dust. Also \citet{grillmair06_5466} investigated the 2d-structure of NGC~5272 with SDSS \textbf{data} and did not detect any tidal structure.

\subsection{NGC~5466}
\begin{figure*}
\centering
\includegraphics[height=9cm]{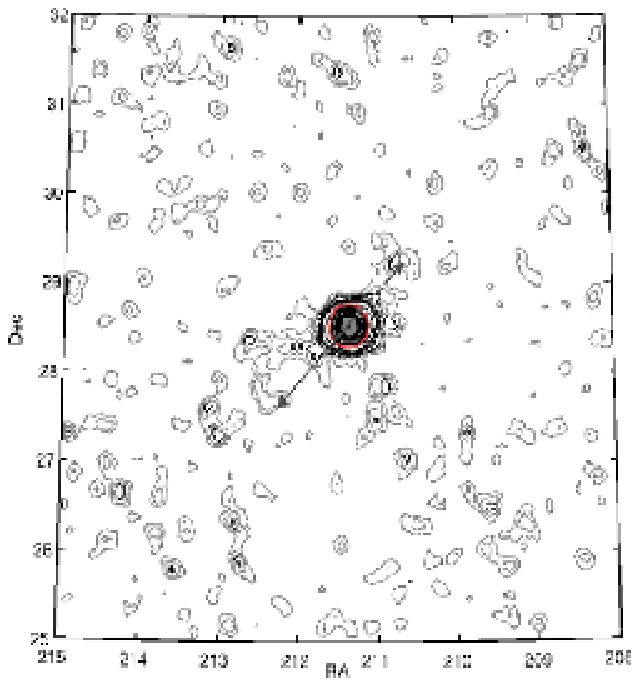}
\includegraphics[height=9cm]{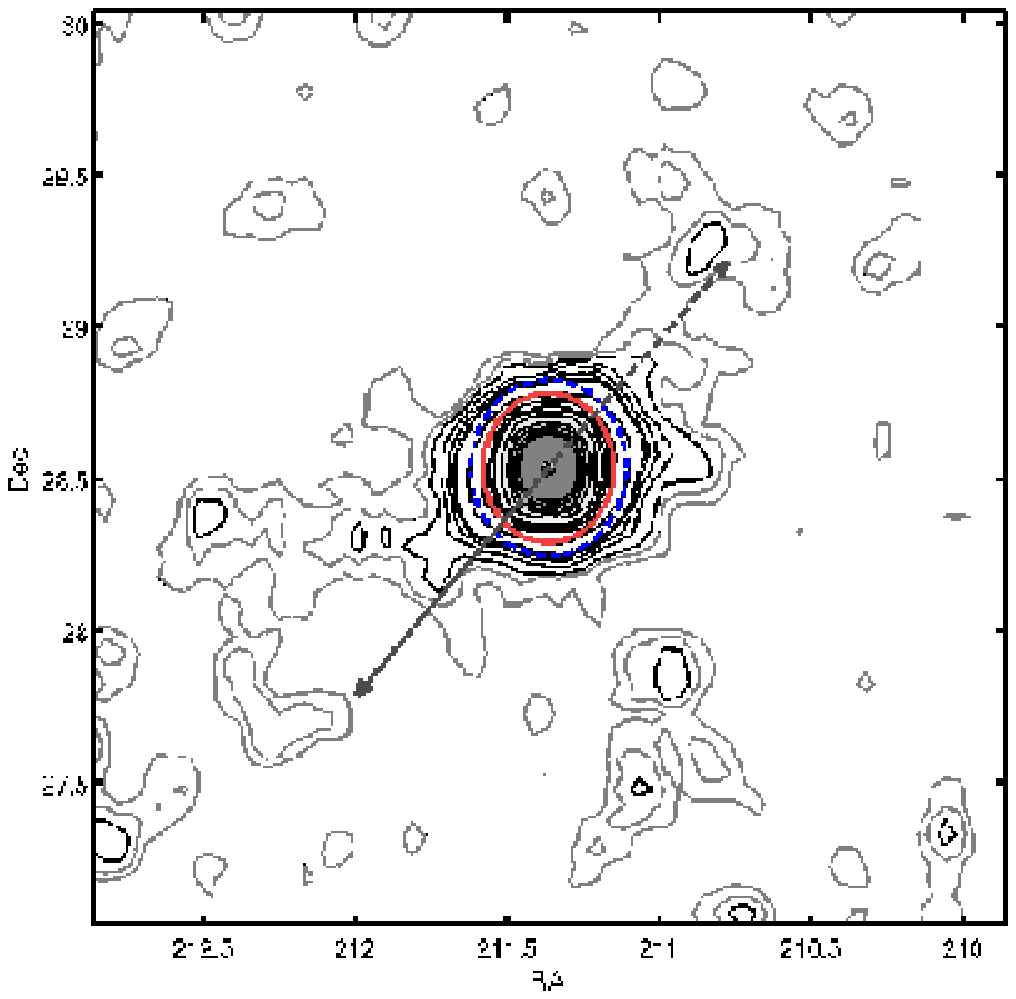}
\caption{Same as Figure~\ref{Fig2419} but for NGC~5466.}
\label{Fig5466}
\end{figure*}

\citet{belokurov06} studied the 2d distribution of NGC~5466 with SDSS data and detected a $4^\circ$-long tidal tail. Earlier studies by \citet{odenkirchen04} and \citet{lehmann97}, on photographic plates, already found indications for the tidal tails of this cluster. \citet{grillmair06_5466} extended this tidal tail to $45^\circ$. Also, \citet{chun10} observed the tidal tails of NGC~5466. We also detect the large extratidal feature of NGC~5466 (Figure~\ref{Fig5466}). The northwest to southeast extension of the tidal tail is clearly visible. The contour plot of the larger area also reveals the tidal tails close to the cluster. The long extensions are hard to distinguish. There are random foreground peaks of up to $5\sigma$ above the mean of $4.5\cdot10^{-2}$~stars arcmin$^{-2}$. This is also the case in the plots of \citet{grillmair06_5466}. Our detections resemble the detections of \citet{chun10}. NGC~5466 is currently at a distance of $15.8$~kpc from the Galactic center, close to its perigalacticon $R_{peri}=6.6$~kpc (D99). The inner tidal tails are aligned with the cluster orbit as predicted by \citet{montuori07}. \citet{fellhauer07} modeled the destruction of NGC~5466 and were able to reproduce the detected tidal tails. They further state that this cluster is mainly destroyed by tidal stripping at each perigalacticon, and that the destruction due to 2-body relaxation plays a minor role. This is in agreement with the destruction rates calculated by GO97.

\subsection{Palomar~5}
\begin{figure*}
\centering
\includegraphics[height=9cm]{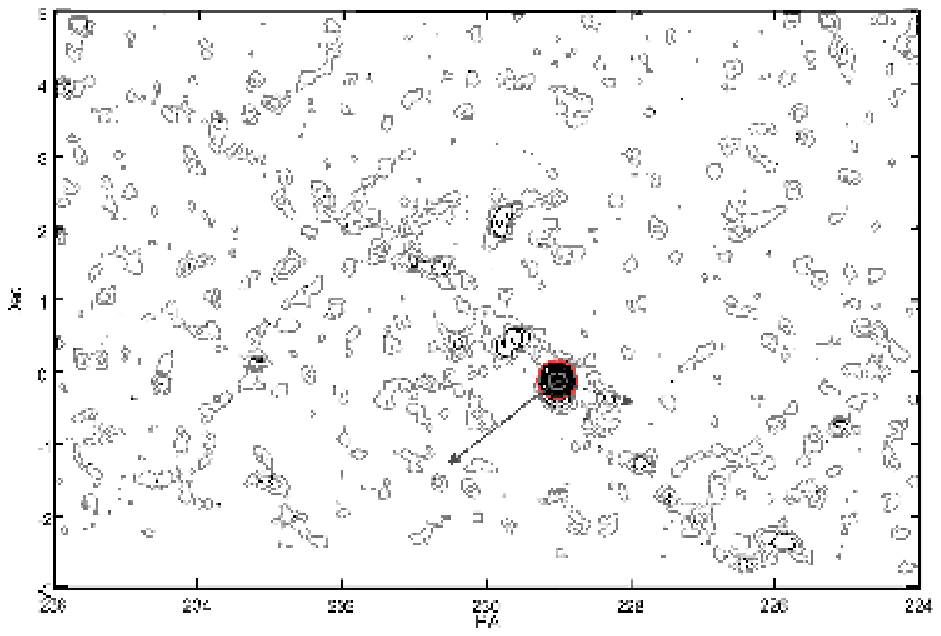}
\includegraphics[height=9cm]{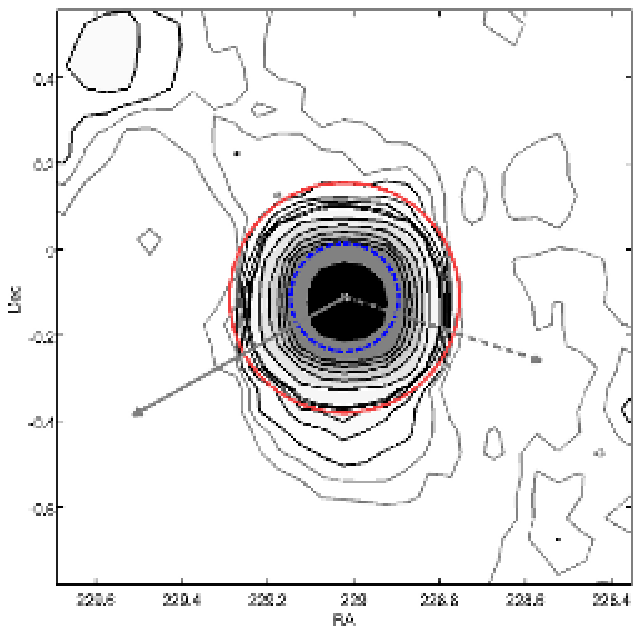}
\caption{Same as Figure~\ref{Fig2419} but for Pal~5. The overdensity at (229.5,2) in the upper panel is the GC NGC~5904.}
\label{FigPal5}
\end{figure*}

Pal~5 is located $18.6$~kpc from the Galactic center. \citet{odenkirchen03} studied the distribution of cluster stars with SDSS data and found tidal tails spanning $10^\circ$ on the sky. Already \citet{odenkirchen01} and \citet{rockosi02} found Pal~5's tidal tails with SDSS data. \citet{grillmair06} extended these tails to $\sim20^\circ$. \citet{odenkirchen02} measured the line-of-sight velocity dispersion of Pal~5 and concluded that the dispersion and the surface density profile are consistent with a King profile of $W_0=2.9$ and $r_t=16.1\arcmin$, similar to the values found here. The disruption of Pal~5 was modeled in \citet{dehnen04} based on the initial observations of \citet{odenkirchen03}. \citet{vivas06} studied RR~Lyrae stars in the QUEST survey. They observed an overdensity of RR~Lyrae stars in the vicinity of Pal~5. Some of these stars trace the cluster's tidal tails. The middle right panel of Figure~\ref{Figprofiles3} shows the number density profile of Pal~5. The overabundance of member stars due to the tidal tails is clearly visible as a pronounced break in the profile. In Figure~\ref{FigPal5} we show the derived contour maps. In the large area view the tidal tails are clearly visible emanating from the cluster. In the zoom-in the $\mathcal{S}$-shape is observed as predicted by theory. The background harbors fluctuations several $\sigma$ above the mean of $1.1\cdot10^{-1}$~stars arcmin$^{-2}$. \citet{koch04} show that the mass segregation in Pal~5 extends to its tidal tails, which contain predominantly low mass stars. Based on a kinematic study of the stars in the tails of Pal~5, \citet{odenkirchen09} suggest that the cluster's orbit is not exactly aligned with its tails.

\subsection{NGC~5904 (M~5)}
\begin{figure*}
\centering
\includegraphics[height=8.5cm]{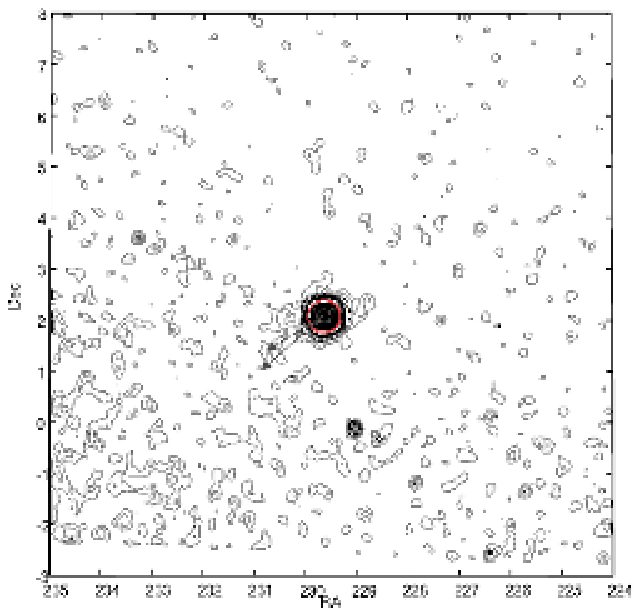}
\includegraphics[height=8.5cm]{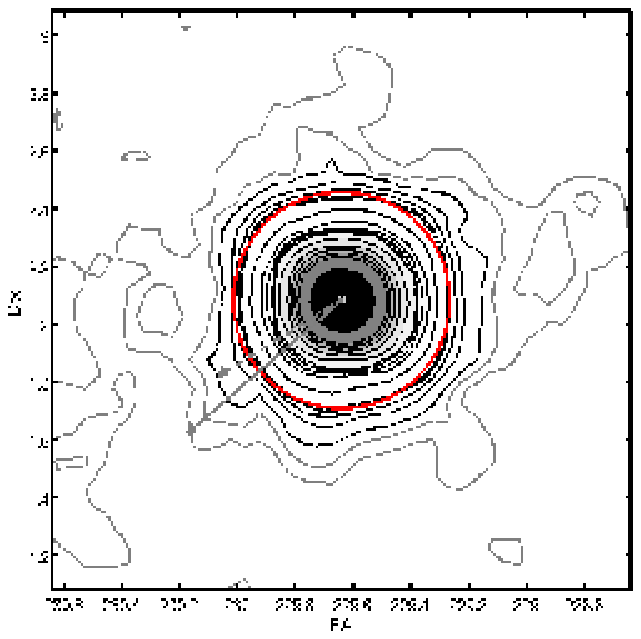}
\caption{Same as Figure~\ref{Fig2419} but for NGC~5904. The overdensity at $(229,0)$ in the left panel is the GC Pal~5.}
\label{Fig5904}
\end{figure*}

NGC~5904 is currently $6.1$~kpc away from the Galactic center, close to its perigalacticon $R_{peri}=2.5$~kpc (D99). I.e., potential tidal tails would be expected to be aligned with the cluster's orbit. In the number density profile, shown in Figure~\ref{Figprofiles2}, no change in slope is observed. In Figure~\ref{Fig5904} we show the contour map. The field background shows contours at levels of $2\sigma$ above the mean of $2.1\cdot10^{-1}$~stars arcmin$^{-2}$. The background density increases towards the Galactic center, but a connection to NGC~5904 does not seem to exist. The tidal tails of Pal~5 are not directly visible, only the main body of this neighboring cluster is detected as an overdensity at $(229,0)$. The three lowest contours of NGC~5904 are distorted, the $1\sigma$-contour is comparable to the background, while the $2$,$3$,$4\sigma$-contours indicate an extratidal halo. The contours are elongated in the direction of motion possibly indicating weak evidence of tidal tails. \citet{leon00} also studied the 2d distribution of this cluster. They also observed round contours, but no extratidal halo. NGC~5904 is approaching perigalacticon, starting its crossing of the Galactic disk \citep{odenkirchen97}.

\subsection{NGC~6205 (M~13)}
\begin{figure*}
\centering
\includegraphics[height=9cm]{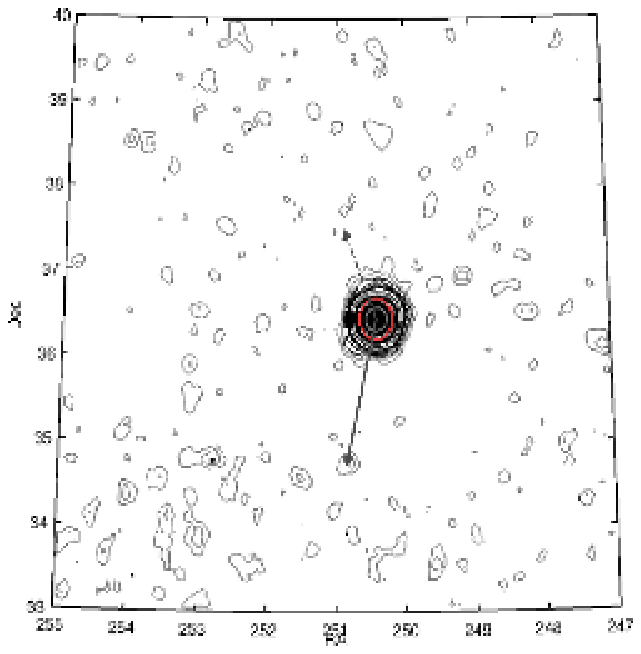}
\includegraphics[height=9cm]{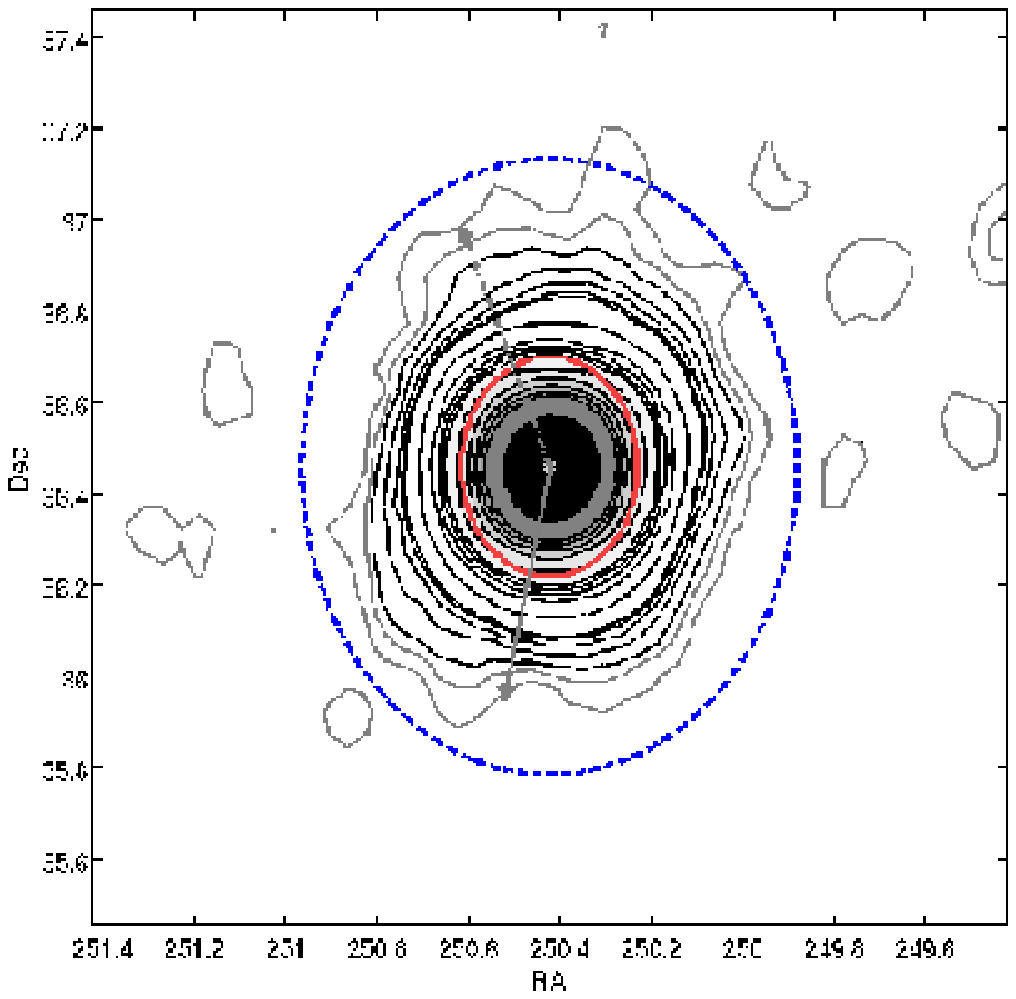}
\caption{Same as Figure~\ref{Fig2419} but for NGC~6205.}
\label{Fig6205}
\end{figure*}

NGC~6205 is currently at a distance of $8.2$~kpc from the Galactic center, close to its perigalacticon $R_{peri}=5.0$~kpc (D99). Thus, we expect potential tidal tails to be aligned with the orbital path. In Figure~\ref{Fig6205} we show the resulting contour plot. The background shows residuals up to $2\sigma$ above the mean of $1.2\cdot10^{-1}$~stars arcmin$^{-2}$. No large scale features are observed. The cluster's central contours are smooth. The contours in the outer regions of the cluster show distortions. We detect a halo of extratidal stars with a slight elongation in the direction of motion. The number density profile in Figure~\ref{Figprofiles2} reveals an overabundance of stars around the tidal radius. \citet{leon00} studied the 2d distribution of NGC~6205 as well. They also detected disturbed contours in the southeast, corresponding to the feature visible in our data. 

\subsection{NGC~6341 (M~92)}
\begin{figure*}
\centering
\includegraphics[height=8.5cm]{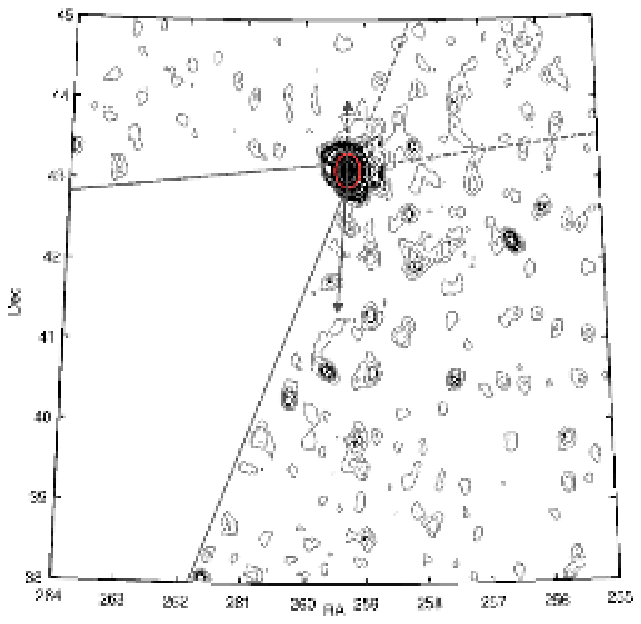}
\includegraphics[height=8.5cm]{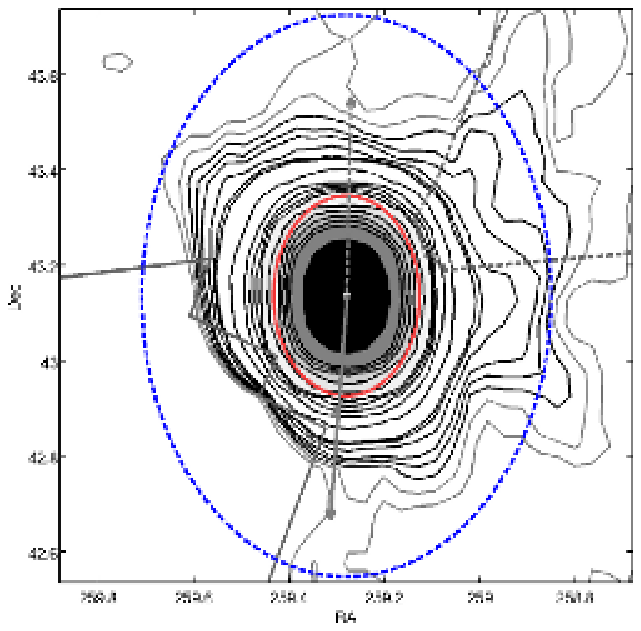}
\caption{Same as Figure~\ref{Fig2419} but for NGC~6341. The solid grey lines denote the edge of the SDSS footprint. The dashed grey line marks the area where the stellar density was doubled (see text for details).}
\label{Fig6341}
\end{figure*}

NGC~6341 is located $8.2$~kpc away from the Sun, close to its apogalacticon, $R_{apo}=9.9$~kpc (D99).  In Figure~\ref{Fig6341} we show our derived contours of NGC~6341. In the left panel the effect of missing data due to the boundary of the SDSS scan regions in the southeast corner is visible. Northeast of the cluster we detected an artifically doubled background density. We corrected for this. All stars in the area denoted by the dashed grey line were studied. If two stars were closer than $2\arcsec$ and similar in color and magnitude one of the stars was deleted from the sample. As seen in the contour map in this way we were able to reach almost the same background level as in any other area of our contour map. The mean background level was determined in the area northeast and southwest of the cluster ignoring the spurious areas. The mean field background is $9.6\cdot10^{-2}$~stars~arcmin$^{-2}$. The background shows overdensities with contours of $3\sigma$ and more mainly southwest of the GC. 

The zoom-in in the lower panel of the same figure reveals mostly smooth contours in the cluster center. The main distortion is in the direction of the corrected background and due to the missing data on the opposite side of the cluster it is not possible to say whether this feature indicates a tidal tail. But also along a southwest to northeast axis we detect extratidal material. The 2d structure of NGC~6341 has been studied before by \citet{testa00}. They observed somewhat elongated contours in the same direction as we do. The number density profile in Figure~\ref{Figprofiles2} shows an overabundance of stars around the tidal radius.

\subsection{NGC~7006}
\begin{figure*}
\centering
\includegraphics[height=9cm]{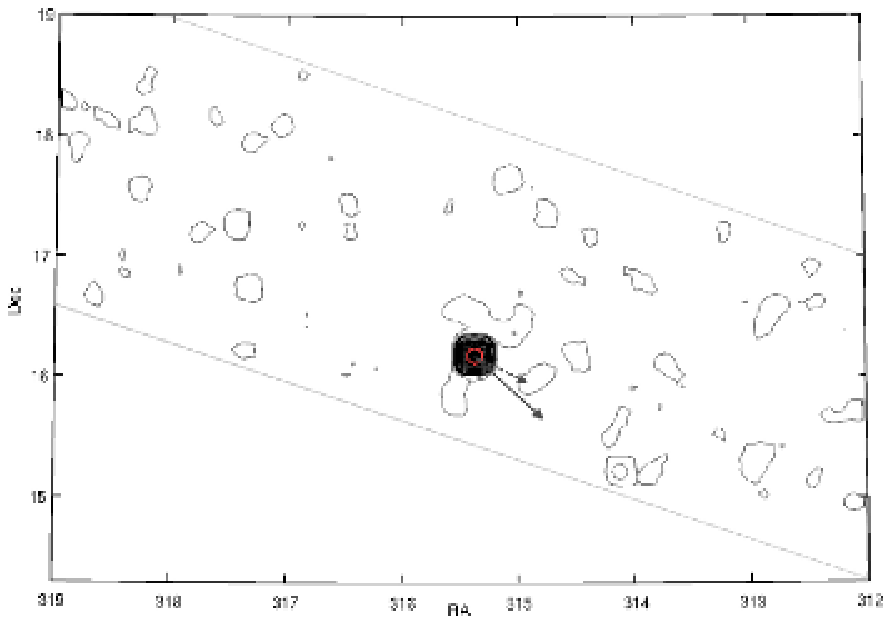}
\includegraphics[height=9cm]{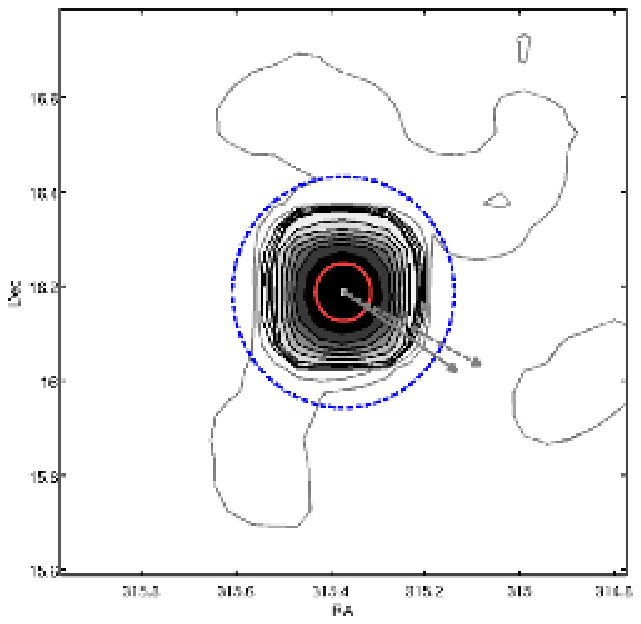}
\caption{Same as Figure~\ref{Fig2419} but for NGC~7006. The solid grey lines denote the edge of the SDSS footprint.}
\label{Fig7006}
\end{figure*}

NGC~7006 is one of the most remote clusters in our sample with a distance to the Galactic center of $R_{MW}=38.8$~kpc. The cluster is currently moving toward the Galactic plane and the Galactic center \citep{dinescu01}. It has not been included in any study on tidal tails searches so far. In Figure~\ref{Fig7006} we show its contour map. We clearly detect the GC as the highest overdensity. The $1\sigma$-contour is very distorted and comparable to the background noise. The mean field background level is $1.2\cdot10^{-1}$~stars arcmin$^{-2}$. The $2\sigma$ and $3\sigma$ contours show some symmetric distortion. All higher contours are smooth, although the halo of extratidal stars for NGC~7006 is huge. For no other cluster in our sample we find such a large extratidal halo.

\subsection{NGC~7078 (M15)}
\begin{figure*}
\centering
\includegraphics[width=\textwidth]{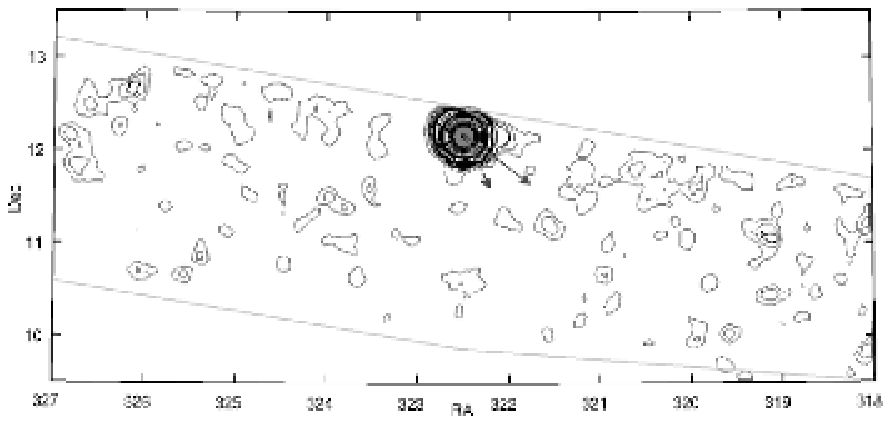}
\includegraphics[height=9cm]{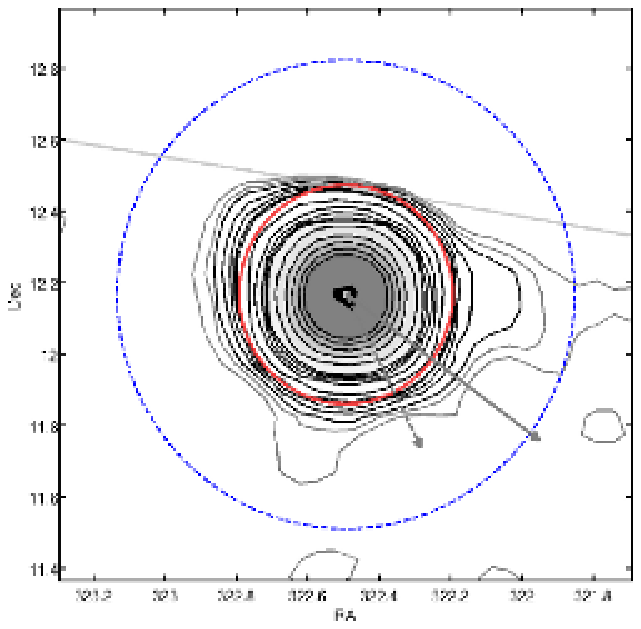}
\caption{Same as Figure~\ref{Fig2419} but for NGC~7078. The solid grey lines denote the edge of the SDSS footprint.}
\label{Fig7078}
\end{figure*}

NGC~7078 is currently located $9.8$~kpc from the Galactic center, close to its apogalacticon $R_{apo}=10.4$~kpc (D99), i.e., we do not expect any correlation between the orbit and the (possible) tail's orientation. In Figure~\ref{Fig7078} we show the contour map. The field around the cluster is very smooth. It shows two peaks of $3\sigma$ above the mean of $3.0\cdot10^{-1}$~stars arcmin$^{-2}$, but not in any symmetric configuration around the cluster. No large scale structures are detected. The contours at the cluster center are undisturbed, but the contours at the tidal radius show some distortions. The proximity of NGC~7078 to the edge of the SDSS survey introduces some asymmetry. Nevertheless, we observe an extratidal extension in southwestern direction. \citet{grillmair95} studied the 2d distribution of NGC~7078. They found an excess of cluster stars extending toward the southeast and a further density peak just north of the cluster. These features do not show up in our data. Instead, we confirm the features seen in the immediate surroundings of NGC~7078 that were also found by \citet{chun10}. Owing to the borders of the SDSS survey, their northwestern feature is not covered by our data. On a larger scale, outside the field of view of \citet{chun10}, we see extended low-density ($1-2\sigma$) structures in roughly east-west direction. These are not aligned with the direction toward the Galactic center, but appear at an angle of approximately $20-30\deg$ (Figure~\ref{Fig7078}), suspiciously aligned with the direction of the SDSS scan. While suggestive, deeper data are needed to confirm or rule out whether these are extended tidal tails. We observe a similar extratidal structure of the contour lines. Figure~\ref{Figprofiles2} shows the number density profile of NGC~7078. We detect a change in the  profile's slope.

\subsection{NGC~7089 (M2)}
\begin{figure*}
\centering
\includegraphics[width=\textwidth]{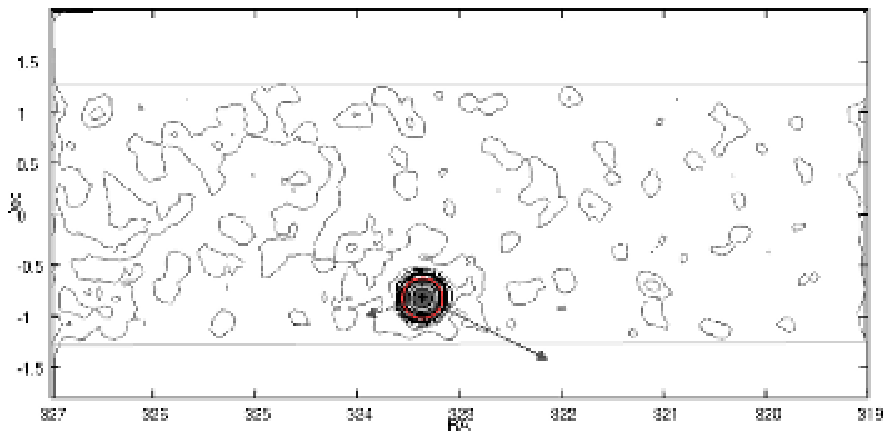}
\includegraphics[height=9cm]{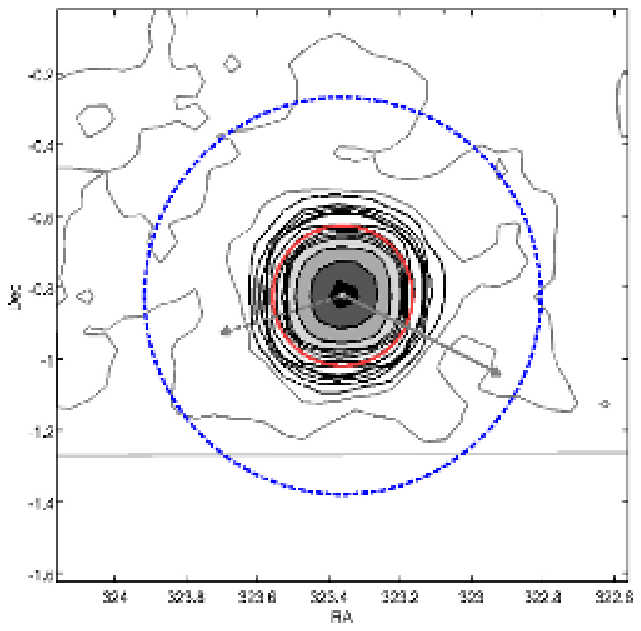}
\caption{Same as Figure~\ref{Fig2419} but for NGC~7089. The solid grey lines denote the edge of the SDSS footprint.}
\label{Fig7089}
\end{figure*}

NGC~7089 is located $10.1$~kpc from the Galactic center, close to perigalacticon $R_{peri}=6.4$~kpc (D99). I.e. a possible inner tidal tail is expected to be aligned with the orbital path of the cluster. In Figure~\ref{Fig7089} we show the contour plot of NGC~7089. The inner contours are smooth. The extended $1\sigma$-contour is comparable to the background contours. The overall background shows at most contours of $2\sigma$. The mean background level is at $1.2\cdot10^{-1}$~stars arcmin$^{-2}$. No large scale features are detected. NGC~7089 shows an extratidal halo, which is a spherical structure. No $\mathcal{S}$-shape or elongation is observed. Also \citet{grillmair95} studied the 2d structure of NGC~7089. They found an excess of cluster stars extending along an E-W axis. In our contours only the $2\sigma$-contours might show some elongation, but overall the halo is quite circular.

\citet{dalessandro09} derived a number density profile and also detected a slight excess of stars beyond the tidal radius compared to the best fit King model. Our number density profile is shown in the Figure~\ref{Figprofiles2}. We detect a change in slope around the tidal radius as expected for a halo of extratidal stars.

\subsection{Palomar~1}
\begin{figure*}
\centering
\includegraphics[height=8.5cm]{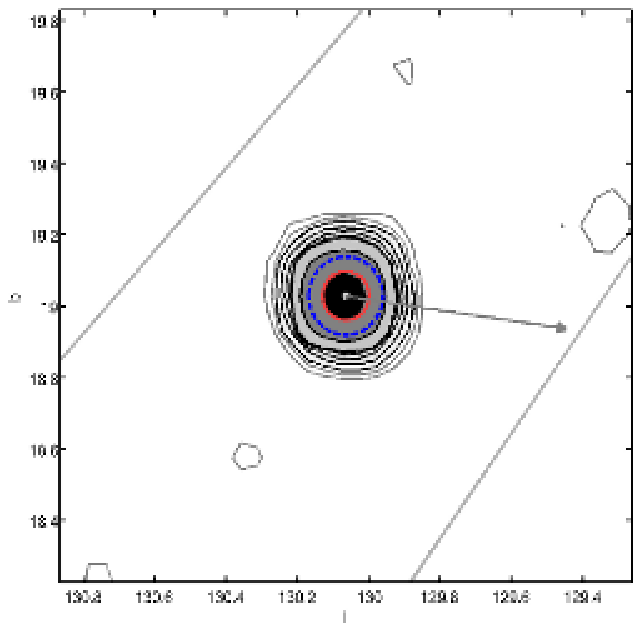}
\caption{Same as Figure~\ref{Fig2419} but for Pal~1. The coordinates shown here are Galactic longitude and latitude.}
\label{FigPal1}
\end{figure*}

Pal~1 is not a typical GC. Its CMD is quite unusual. It has a populated main sequence, but hardly any stars on the red giant and the horizontal branch. This feature was noticed by \citet{borissova95}, \citet{rosenberg98}, and \citet{sarajedini07}. Further, this cluster might be up to $8$~Gyr younger(!) than a typical GC like 47~Tuc \citep{sarajedini07}. It is also discussed in the literature, whether Pal~1 is a member of the Monoceros stream/Canis Major dwarf \citep{crane03, forbes10}. It cannot be ruled out that Pal~1 is misclassified as a GC since objects of this young an age are usually considered open clusters. The low background density  contour map shown in Figure~\ref{FigPal1} is due to the cluster's high declination plotted in Galactic longitude and latitude, ($l$, $b$). Although Pal~1 has a high destruction probability (GO97), mainly driven by disk and bulge shocks, it does not show any distortions of its contours, except for a large circular halo of extratidal stars (comparable to NGC~7006). In the field we see no density peaks. The number density profile is shown in Figure~\ref{Figprofiles3}. The profile drops outside of $\sim 30$~arcmin because the edge of the survey is reached. In summary we can say that we detect a symmetric overdensity at the position of Pal~1.

\subsection{Palomar~3}
\begin{figure*}
\centering
\includegraphics[height=8.5cm]{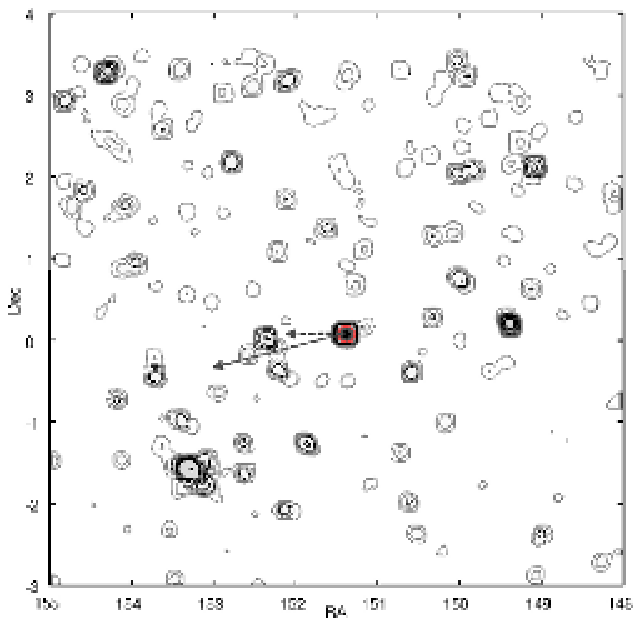}
\includegraphics[height=8.5cm]{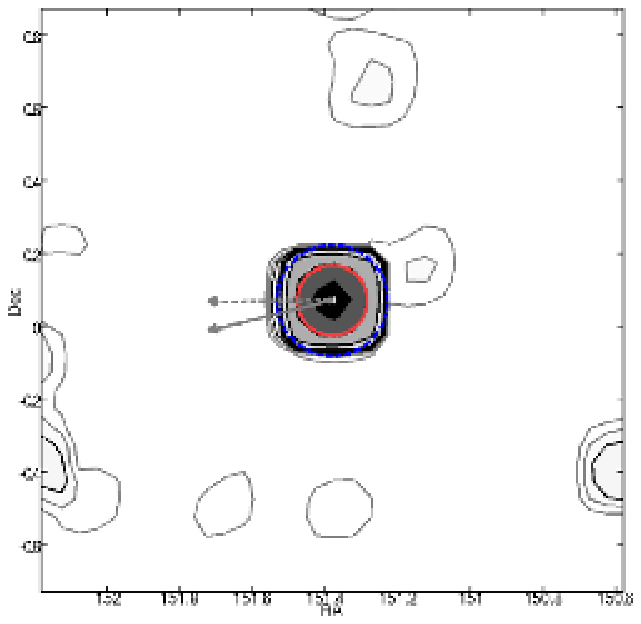}
\caption{Same as Figure~\ref{Fig2419} but for Pal~3.}
\label{FigPal3}
\end{figure*}

Pal~3 is currently at a distance of $R_{MW}=84.9$~kpc, very close to perigalacticon $R_{peri}=82.5$~kpc (D99). As this is the cluster's closest approach to the Galaxy we do not expect to see any large signs of dissolution due to tidal shocks. Also the destruction rates derived by GO97 are small. In Figure~\ref{FigPal3} we show the contour map of a large area centered on Pal~3. The cluster can clearly be detected as the strongest overdensity. Nevertheless, the background shows several pronounced overdensities. The largest overdensity at ($153.5$,$-1.5$) is the signal of the dwarf spheroidal galaxy Sextans. The cluster's contours look undisturbed. We observe a distortion of the contours towards the Galactic center, but no corresponding feature on the opposite side. Generally, Pal~3 has a large extratidal halo which shows a sharp edge. The number density profile is shown in Figure~\ref{Figprofiles3}. We note that owing to the sparseness of Pal~3's CMD and its large distance deeper data are needed to investigate options such as a connection with Sextans.

\subsection{Palomar~4}
\begin{figure*}
\centering
\includegraphics[height=8.3cm]{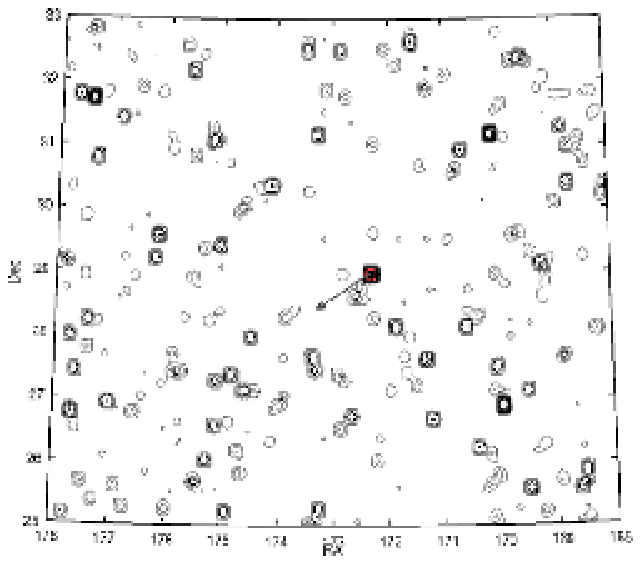}
\includegraphics[height=8.3cm]{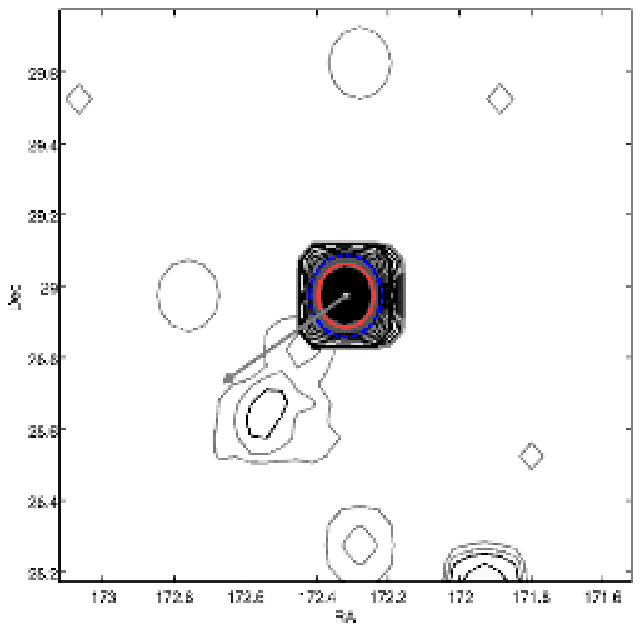}
\caption{Same as Figure~\ref{Fig2419} but for Pal~4.}
\label{FigPal4}
\end{figure*}

A density plot of Pal~4 and surroundings is shown in Figure~\ref{FigPal4}. We discover a large halo of extratidal stars. The $1\sigma$-contour is stretching along towards the Galactic center, but only in one direction. The field around Pal~4 is showing strong overdensitie of up to $10\sigma$. These density peaks show steep edges and in the center broad plateaus. It is unclear where these features are coming from. Comparing these field peaks with the cluster's tidal tail, hardly any difference can be seen. An interesting structure of the contour lines is visible at the cluster's western edge.

\citet{sohn03} studied the 2d-morphology of Pal~3 and Pal~4 using CFH12K wide-field photometry. For Pal~3 they detect some extensions in the direction towards the Galactic center and anti-center which are not more prominent than $1\sigma$ over the background. For Pal~4 they see a tail extending to the NW. In our data we do not observe such a feature. Pal~4 is the most distant cluster in our sample. Again, deeper data are needed to reveal the possible existence of tidal tails.

\subsection{Palomar~14}
\begin{figure*}
\centering
\includegraphics[height=8.5cm]{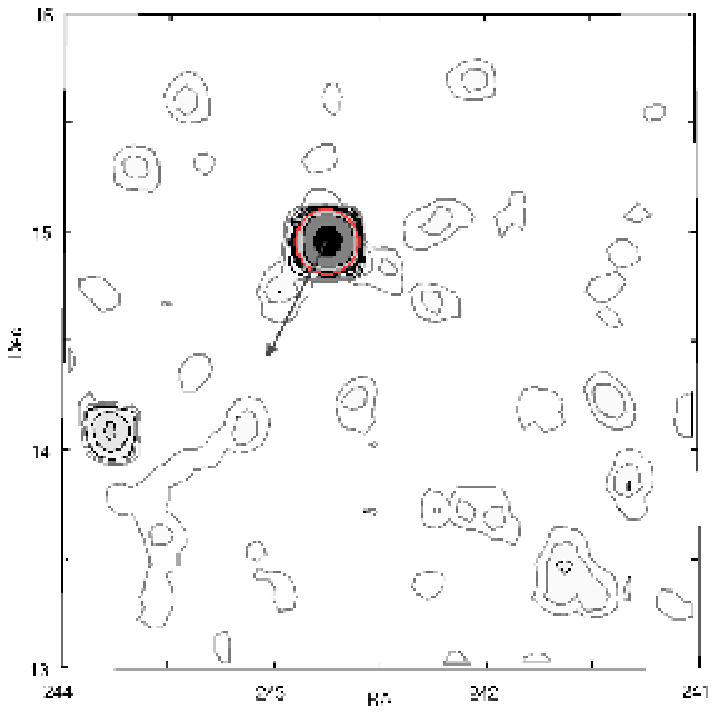}
\includegraphics[height=8.5cm]{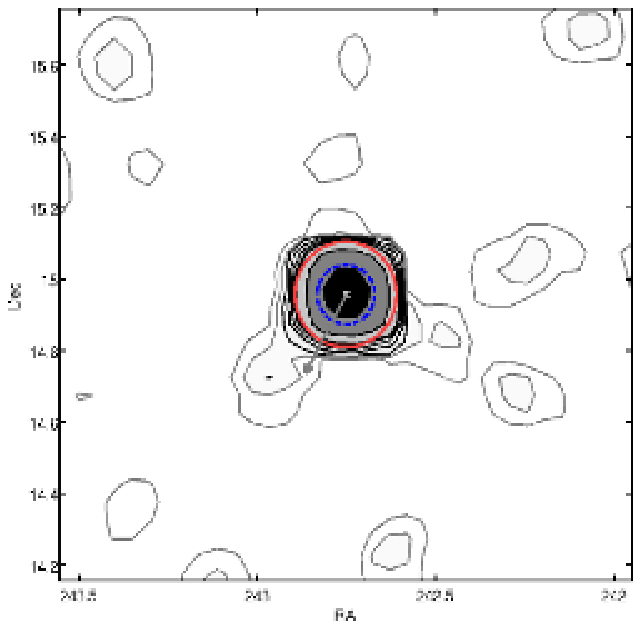}
\caption{Same as Figure~\ref{Fig2419} but for Pal~14.}
\label{FigPal14}
\end{figure*}

In Figure~\ref{FigPal14} we show the contour map of Pal~14. \citet{martinez04} studied the CMD of Pal~14 and compared it to the CMD of a region outside the cluster. They detected some hints for extratidal stars in a region of the CMD. We see smooth central contours and an extratidal halo. The field background shows some overdensities of up to $5\sigma$ above the mean which is $1.47\cdot10^{-1}$~stars arcmin$^{-2}$. The $1\sigma$ and $2\sigma$ contours show an elongated distortion, comparable to the field background contours. The extratidal halo is spherical with some asymmetry towards the southeast. Whether this indicates the beginning of sparsely populated tidal tails remains unclear. Also for this GC, deeper data would be desirable in order to confirm this or to rule it out.

\section{Discussion}
\label{Secdiscussion}

\subsection{Horizontal branch stars}
From theoretical investigations and simulations \citep[e.g.,][]{baumgardt03} we know that preferentially stars of lower mass are lost. Measurements of the mass function of Pal~5 and its tidal tails confirm this picture \citep{koch04}. Also the number density of main-sequence stars is larger than for RGB stars. Therefore, tracing tidal structures with only upper RGB stars may be difficult, as is the case for Pal~3, Pal~4, Pal~14 and NGC~2419 and this es exacerbated by the sparsely populated RGBs of the Palomar GCs. Nonetheless, we do find halos of extratidal stars as possible signs for some dissolution for all four clusters. But the field background is noisy and shows overdensities of more than $3\sigma$.

The blue horizontal branch (BHB) and RR~Lyrae stars are located at a position in the CMD where hardly any field stars are located. Therefore any BHB or RR~Lyrae star at the cluster distance found outside the cluster's tidal radius is with a high probability a former cluster member. For NGC~5466, \citet{belokurov06} found that the BHB/RR~Lyrae stars also trace the tidal tails. For those of our GCs that show no large tidal structures, but only distorted tidal contours the majority of the BHB/RR~Lyrae stars are inside the tidal radius. As an example: In NGC~6341, we found 194 BHB/RR~Lyrae stars. Only 20 of those are found outside the tidal radius. The remaining 174 BHB/RR~Lyrae stars are all found within the tidal radius (red ellipse in Figure~\ref{Fig6341}). For NGC~4147, for which we have found a complex multiple arm morphology and extended $1\sigma$ structures that look like weak tidal tails, most BHB stars are within the innermost drawn contour, and $30\%$ are found outside. The BHB stars do not trace the multiple arms. But the BHB/RR~Lyrae stars in the field are mostly found in overdensities. E.g., at (181.2, 17) two overdensities of $9\sigma$ and $10\sigma$, respectively are found. In the center of both a single BHB/RR~Lyrae star is detected. At the same time, there are BHB stars that are not connected to any overdensity. The algorithm used to derive the density of cluster member stars gives high weights to BHB and RR~Lyrae stars. Without spectroscopic information it is not possible to draw any reliable conclusion whether BHB/RR~Lyrae stars connected to a prominent overdensity in the background are `lost' cluster members. But the existence of BHB/RR~Lyrae stars without any overdensity is supporting the hypothesis that the stars in those overdensities are more likely former member stars.

A similar connection between some of the cluster's BHB/RR~Lyrae stars and overdensities in the field is observed for NGC~5024, NGC~5053, NGC~5466, NGC5904, NGC~6205, and NGC~7006. Our current data do not permit us to conclude whether these overdensities are related to the GCs.

For some GCs in our sample a connection to the Sagittarius stream \citep[e.g.,][]{bellazzini03b,law10,forbes10} or the Monoceros stream \citep{crane03,forbes10} is discussed in the literature. It is possible that the detected extratidal structures and overdensities in the field of NGC~4147, NGC~5024, NGC~5053 and Pal~1 are features induced by the field populations of the disrupted dwarf galaxies. The data used here do not allow us to make a definite conclusion.

\subsection{Comparison to theory}
The vital diagram of GO97 in Figure~\ref{figVital} predicts a dissolution of NGC~4147 due to 2-body relaxation. The multiple arm feature observed in Figure~\ref{Fig4147} for NGC~4147 supports the cluster's dissolution. Furthermore, the simulations by \citet{montuori07} found such a multiple arm morphology for GCs on eccentric orbits approaching apocenter. NGC~4147's current position close to apogalacticon matches the simulations well. Of all stars in the distorted $1\sigma$-contour around NGC~4147 about $13\%$ are found outside the $4\sigma$-contour. The $4\sigma$-contour corresponds roughly to the Jacobi radius of NGC~4147 \citep{baumgardt09}, i.e., stars outside the $4\sigma$-contour left the potential field of the cluster. The tidal features of NGC~4147 are small compared to the ones for Pal~5. Pal~5 is close to destruction due to tidal shocks, while NGC~4147's dissolution is mainly driven by relaxation.

The Palomar clusters in our sample also lie outside the triangles in Figure~\ref{figVital}. As the outermost triangle is only appropriate for clusters with a current distance to the Galactic center smaller than $12$~kpc, and Pal~3, Pal~4, and Pal~14 are currently further than $69$~kpc away from the Milky Way no conclusions can be drawn from the diagram for these three clusters. Pal~5 is known to have tidal tails. Its destruction rate was studied in detail by \citet{dehnen04}.

Pal~1 is also located outside the triangles in the vital diagram. Hence a dissolution due to 2-body relaxation is likely, but the contour map in Figure~\ref{FigPal1} shows no typical $\mathcal{S}$-shape. It reveals a large, spherical  extratidal halo though. The halo extends much further out than the cluster's Jacobi radius. The Jacobi radius is slightly smaller than the $20\sigma$-contour (the second most inner contour). $71\%$ of all stars located within the $1\sigma$-contour are found outside the Jacobi radius, i.e., should not be bound to the GC anymore. Pal~1 in general is an interesting object. Its CMD shows a clear main-sequence but a very sparsely populated red giant branch. The low stellar density makes it difficult to detect tidal tails if present. The question of whether Pal~1 is really a GC or rather an open cluster remains unsolved.

The remaining GCs in our sample from the NGC catalog are all located well within the triangles of GO97's vital diagram. Hence, no (pronounced) signs for dissolution are expected.  NGC~5466 is an exception and shows that the theoretical calculations can be improved, as tidal tails have been detected. For NGC~5053 some contours are located outside the Jacobi radius, a possible sign for unbound cluster stars in the northeast, west, and southwest. Generally the clusters show an extratidal halo, which is to be expected from dynamical studies. These GCs are not in danger to be destroyed within the next Hubble time, according to GO97 and D99. This is supported by the lack of large scale tidal features in our study.

\subsection{Number density profiles and halos of extratidal stars}
Most of the clusters in our sample show a (pronounced) halo of stars outside the tidal boundary. \citet{combes99} simulated GCs in a tidal field and observed such a halo as well. The stars outside the tidal radius spread out in density like a power law. \citet{combes99} transformed the 3-dimensional simulations into a (observable) projection on the sky and derived the power law to be ``$r^{-3}$ or steeper''. The extratidal halos observed by \citet{grillmair95} and \citet{leon00} show only slopes of $-1$. \citet{combes99} argue that the discrepancy is a consequence of noisy background-foreground subtraction. Our stellar sample contains very little contamination. Therefore, an investigation of these halos is of interest. Furthermore, \citet{grillmair95} observed for clusters with tidal extensions a break in their number density profiles, becoming pure power laws at larger radii. 

\citet{leon00} introduced $Q^b_a$ as the slope of the number density profile between $a\cdot r_t$ and $b\cdot r_t$. They chose to compute the three slopes: $Q^3_1$, $Q^6_3$, and $Q^6_1$. We applied the same method to the clusters in our sample and derived the three slopes. In the profile shown in Figure~\ref{Figprofiles1}-\ref{Figprofiles3} slope changes are visible in many clusters. The slope $Q^6_3$ was only computable for clusters, where the background was not too noisy. In order to fit the power law slopes we performed a least-squares fit between the two radii. In Table~\ref{tabslopes} we list the resulting slopes for all clusters. 

The theoretical prediction of slopes of ``$r^{-3}$ or steeper'' is observed for many clusters in our sample for $Q^3_1$. Only NGC~4147, NGC~5053, Pal~4, and Pal~5 have a significantly flatter slope than predicted. The measurement for Pal~3 is affected by a large error. The fits for $Q^6_3$ show large errors or were in some cases not possible at all, due to the noisy background. For Pal~1, we did not perform any fit due to the strange shape of the number density profile. In general, the simulated halos of \citet{combes99} reproduce well the observed halos.
\begin{table}
\centering
\caption{Power law slopes $Q^b_a$ of the radial density profiles between the two radii $a\cdot r_t$ and $b\cdot r_t$. The stated errors are only the errors from the fit.\label{tabslopes}}
\begin{tabular}{llccc}
\hline
\hline
\noalign{\smallskip}  
{NGC}& Name  &$-Q^3_1$&  {$-Q^6_1$} &  {$-Q^6_3$}  \\
\noalign{\smallskip}     
\hline
2419 &       &$3.44\pm0.75$&$3.32\pm0.47$ &$1.13\pm2.4$\\
4147 &       &$1.48\pm0.24$ &$1.70\pm0.24$&$4.19\pm0.32$ \\
5024 & M~53  &$3.73\pm0.35$&$3.69\pm0.24$&$1.68\pm1.28$ \\
5053 &       &$0.33\pm1.05$&$2.35\pm1.76$&\dots \\
5272 & M~3   &$0.94\pm0.38$&$0.91\pm0.16$&\dots \\
5466 &       &$3.89\pm0.46$&$1.52\pm4.01$&$3.39\pm0.75$ \\
5904 & M~5   &$3.33\pm3.72$&$3.33\pm0.24$&$-0.70\pm1.83$ \\
6205 & M~13  &$3.28\pm0.33$&$3.28\pm0.20$&$1.65\pm6.80$  \\
6341 & M~92  &$3.65\pm0.21$&$3.76\pm0.21$&$3.76\pm11.0$ \\
7006 &       &$3.38\pm0.73$&$3.36\pm0.51$&$1.23\pm6.38$ \\
7078 & M~15  &$2.88\pm0.26$&$3.00\pm0.21$&\dots\\
7089 & M~2   &$2.58\pm0.21$&$2.65\pm0.15$&$5.09\pm2.58$ \\
     & Pal~3 &$7.14\pm10.58$&$7.16\pm8.27$&\dots\\
     & Pal~4 &$1.47\pm1.15$&$1.88\pm0.77$&\dots\\
     & Pal~5 &$0.26\pm0.42$&$1.33\pm0.29$&$3.48\pm2.95$ \\
     & Pal~14&$3.13\pm2.96$&$3.46\pm2.58$&\dots \\
\hline
\end{tabular}
\end{table}

For most GCs an extratidal halo exists, while  the Jacobi radius is larger than this halo. This is the case for NGC~2419, NGC~5024, NGC~5272, NGC~5904, NGC~6205, NGC~6341, NGC~7006, NGC~7078, and NGC~7089. From the definition of the Jacobi radius this suggest that the stars in the extratidal halo have not left the cluster potential. The derivation of the Jacobi radius is based on measurements and assumptions, e.g., the mass-to-light ratio is assumed to be 2, which may not be generally true, see M05 for mass-to-light ratios of the Galactic GCs. One possibility is that the contour maps do not reveal halos of extratidal stars but just the halo of the GCs. Although, the simulations by \citet{combes99} reveal extratidal halos with the identical power-law slope as our observed halos.
 
The Palomar clusters have a halo of stars outside the Jacobi radius. \citet{baumgardt09} classifies these clusters as ``tidally filled''. We actually observe that these clusters are even more extended. Pal~14 is a very special case: the Jacobi radius is smaller than the tidal radius from the King profile. The low measured velocity dispersion of Pal~14 does not suggest a dissolved GC \citep{jordi09}, neither do the mostly symmetrical contour lines. For Pal~5 the same observation is made, the Jacobi radius is smaller than the tidal radius. Pal~5  has a lower than expected velocity dispersion, although it will be destroyed within the next 110~Myr \citep{odenkirchen02}. It is possible that the large distance to Pal~14 prevents us from detecting its tidal tails although we do see possible extensions at the $1-2\sigma$ level. Or their orientation makes them difficult to detect. Simulations of dwarf spheroidals have shown that the detection of the characteristic $\mathcal{S}$-shape of tidal tails depends on the position of the orbit relative to the line of sight \citep{munoz08}.
\section{Summary \& Conclusions}
\label{Secsummary}
We studied the 2d-distribution of (potential) member stars on the sky of the 17 GCs in the SDSS. The stars were chosen from the SDSS DR~7 and from An08. All stars in a field around a GC were selected in color and magnitude to match the observed colors and magnitudes of cluster stars.

We derived the number density profiles for the 17 clusters in our sample. Thanks to the large area coverage of the SDSS, we were able to trace the profiles as far out as to the flat field star background. In some cases a break in the profile was detected and interpreted as a sign for an extratidal halo. Comparing the profiles with the 2d-distribution of cluster stars on the sky, we detected a halo of apparent extratidal stars for more clusters than a clear break in the profile. 

A color-magnitude weighted counting algorithm from \citet{odenkirchen03} was used to derive density maps of cluster stars on the sky. We recovered the known extended tidal tails of Pal~5 and NGC~5466. For NGC~4147 we detected a complex, \textbf{two} arm feature of extratidal stars, and for NGC~5904 we find a roughly symmetric extension that might indicate short and stubby tails. For NGC~7006 and Pal~1 the extratidal halo is remarkably large. For Pal~1, $71\%$ of the stars of the GCs are found outside the Jacobi radius, i.e., are not bound to the cluster anymore. Nevertheless, they are arranged spherically around the cluster. No sign of tidal effects is observed. We also detected  extratidal features for NGC~5053, comparable to the findings of \citet{lauchner06}. We confirm extratidal features in the immediate surroundings of NGC~7078 \citep{chun10}. For Pal~14 deeper data are needed in order to investigate whether the weakly visible extensions constitute tidal tails.

The comparison of our observations to the theoretical study of the GCs' destruction rates by GO97 shows generally good agreement. Most NGC clusters in our sample were not expected to show signs for dissolution. They mostly only revealed a halo of extratidal stars well within their Jacobi radius. These extratidal halos show for the majority of our sample the theoretically expected ``$r^{-3}$'' density slope.

\begin{acknowledgements}
KJ would like to thank Shoko Jin for her help in deriving the direction of motion for each GC. We thank an anonymous referee for useful comments. KJ and EKG acknowledge support from the Swiss National Foundation through grant numbers 20020-122140 and 20020-113697. 

Funding for the SDSS and SDSS-II was provided by the Alfred P. Sloan Foundation, the Participating Institutions, the National Science Foundation, the U.S. Department of Energy, the National Aeronautics and Space Administration, the Japanese Monbukagakusho, the Max Planck Society, and the Higher Education Funding Council for England. The SDSS was managed by the Astrophysical Research Consortium for the Participating Institutions. The Participating Institutions are American Museum of Natural History, Astrophysical Institute Potsdam, SEGUE and Supernovae, University of Basel, Cambridge University, Case Western Reserve University, University of Chicago, Drexel University, Fermi National Accelerator Laboratory, Institute for Advanced Study, Princeton Legacy, Japan Participation Group, Johns Hopkins University, Joint Institute for Nuclear Astrophysics, Kavli Institute for Particle Astrophysics and Cosmology, Korean Scientist Group, LAMOST, Los Alamos National Laboratory , Max-Planck-Institute for Astronomy/Heidelberg, Max-Planck-Institute for Astrophysics/Garching, New Mexico State University, Ohio State University, University of Pittsburgh, University of Portsmouth, Princeton University, US Naval Observatory, University of Washington.
\end{acknowledgements}


\end{document}